\documentclass{article}

\usepackage{PRIMEarxiv}

\usepackage[utf8]{inputenc} 
\usepackage[T1]{fontenc}    
\usepackage{hyperref}       
\usepackage{url}            
\usepackage{booktabs}       
\usepackage{amsfonts}       
\usepackage{nicefrac}       
\usepackage{microtype}      
\usepackage{lipsum}
\usepackage{fancyhdr}       
\usepackage{graphicx}       
\graphicspath{{media/}}     
\usepackage{setspace}
\usepackage{indentfirst}
\usepackage{float}
\usepackage{caption}
\usepackage[belowskip=-10pt,aboveskip=5pt]{caption}
\usepackage{subcaption}
\usepackage{relsize}
\usepackage{mathrsfs,amsmath}
\usepackage{mathtools}
\usepackage{algorithm}
\usepackage{algpseudocode} 
\usepackage{color,soul}

\usepackage{tikz}
\usetikzlibrary{arrows,calc,fit}
\tikzset{box/.style={draw, rectangle, thick, node distance=10em, text width=8.5em, text centered, minimum height=1.em}}
\tikzset{container/.style={draw, rectangle, dashed, inner sep=1em}}
\tikzset{line/.style={draw, thick, -latex'}}

\hypersetup{
	colorlinks=true,
	linkcolor=blue,
	citecolor=blue,
	urlcolor=blue,
	pdftitle={Multi-objective Optimization under Uncertainty of Part Quality in Fused Filament Fabrication},
	pdfkeywords={additive manufacturing, material extrusion, fused filament fabrication, robust design optimization, multi-objective optimization, optimization under uncertainty, process design, Bayesian neural network, machine learning},  
	pdfauthor={Berkcan Kapusuzoglu, Paromita Nath, Matthew Sato, Sankaran Mahadevan, Paul Witherell},
}

\title{Multi-Objective Optimization Under Uncertainty of Part Quality in Fused Filament Fabrication
\thanks{\textit{\underline{Citation}}: 
\textbf{Kapusuzoglu, B., Nath, P., Sato, M., Mahadevan, S., \& Witherell, P. Multi-objective optimization under uncertainty of part quality in fused filament fabrication. \textit{ASCE-ASME Journal of Risk and Uncertainty in Engineering Systems, Part B: Mechanical Engineering} 8(1), 011112 (2022). DOI: \href{https://doi.org/10.1115/1.4053351}{10.1115/1.4053351}}
} 
}

\author{
  \textbf{Berkcan Kapusuzoglu}\thanks{Corresponding author: berkcan.kapusuzoglu@vanderbilt.edu}$^{\ ,1}$, \textbf{Paromita Nath}$^1$, \textbf{Matthew Sato}$^1$, \\ \textbf{Sankaran Mahadevan}$^1$, \textbf{Paul Witherell}$^2$ \\
  \vspace{0.2cm} \\
  $^1$Department of Civil and Environmental Engineering, Vanderbilt University, Nashville, TN 37235, USA \\
  $^2$Systems Integration Division, Engineering Laboratory, \\ National Institute of Standards and Technology (NIST), Gaithersburg, MD 20899, USA
}

\begin{document}
\maketitle

\begin{abstract}
This work presents a data-driven methodology for multi-objective optimization under uncertainty of process parameters in the fused filament fabrication (FFF) process. The proposed approach optimizes the process parameters with the objectives of minimizing the geometric inaccuracy and maximizing the filament bond quality of the manufactured part. First, experiments are conducted to collect data pertaining to the part quality. Then, Bayesian neural network (BNN) models are constructed to predict the geometric inaccuracy and bond quality as functions of the process parameters. The BNN model captures the model uncertainty caused by the lack of knowledge about model parameters (neuron weights) and the input variability due to the intrinsic randomness in the input parameters. Using the stochastic predictions from these models, different robustness-based design optimization formulations are investigated, wherein process parameters such as nozzle temperature, nozzle speed, and layer thickness are optimized under uncertainty for different multi-objective scenarios. Epistemic uncertainty in the prediction model and the aleatory uncertainty in the input are considered in the optimization. Finally, Pareto surfaces are constructed to estimate the trade-offs between the objectives. Both the BNN models and the effectiveness of the proposed optimization methodology are validated using actual manufacturing of the parts.
\end{abstract}


\section{Introduction}\label{sec:intro}

Additive manufacturing (AM) is the layer-wise material deposition process to manufacture parts in the desired shape from a computer aided design (CAD) model~\cite{international2015additive}. Although AM is a rapidly expanding technology and has shown immense potential in several industries, application of AM has been limited due to variability in the product quality. Thus, improving the product quality and reducing the variability in AM parts is an active area of research. The traditional trial-and-error approach of optimizing process parameters based on physical experiments is expensive and time-consuming. Therefore, recently there is increasing interest in using model-based methods to study AM and optimize the process parameters. Both physics-based models and data-driven models have been developed to better understand the AM process~\cite{berkcan2020PIML}.

Developing physics-based models is a major direction in AM research. Several physics-based models have been developed depending on the AM process category and the quantity of interest (QoI)~\cite{debroy2017building}. Since AM is a complicated process, a different model may be required to represent a particular sub-stage or phenomenon in the manufacturing process. Multiple physics-based models at different length and time scales are coupled together to simulate the AM process. Sophisticated commercial software are often used to solve the governing equations, and these tend to be computationally demanding. Thus, physics-based modeling of full-scale AM parts has been challenging~\cite{megahed2016metal}.

The potential of data-driven models for quality monitoring and control in AM is also being explored by the AM community~\cite{baumann2018trends}. Data-driven models have been shown to perform well and generally do not require expert, in-depth knowledge of the complex physics of the AM process~\cite{kwon2020convolutional}. Thus, studies have been conducted on using the available experimental data to build data-driven models of the AM process. Khanzadeh et al.~\cite{khanzadeh2018porosity} compared supervised machine learning approaches such as Decision Tree, K-Nearest Neighbor, Support Vector Machine, Linear Discriminant Analysis, and Quadratic Discriminant Analysis to classify melt pools for porosity prediction. With the evolution of data collection techniques and instruments, more and more data is now available in the form of images such as thermal images, optical images, and profilometer data, in addition to the data from instruments such as thermocouples, digital calipers, etc. Meanwhile, with the advancement in computer hardware, researchers also have access to increased computing power. This has led to a growing trend of using artificial neural networks (ANN) as the preferred tool for solving classification and regression problems to handle the large amounts of data.

ANN has been used to predict the geometry of a single bead in wire and arc additive manufacturing (WAAM) from the wire-feed rate and travel speed~\cite{ding2016bead}. An ANN-based prediction model was developed to predict strain recovery rates and transformation temperatures of fabricated metal parts for given laser power, laser speed, and hatch spacing of a selective laser melting (SLM) AM process~\cite{mehrpouya2019prediction}. Ye et al.~\cite{ye2018situ} proposed an adapted deep belief network approach using the plume and spatter signatures obtained during an SLM process to detect defects in the part. Gaikwad et al.~\cite{gaikwad2019situ} built a deep convolutional neural network (CNN) model with in-situ layer-wise optical images as input to predict the quality in terms of statistical features obtained from offline X-ray computed tomography (XCT) of the manufactured part. Kwon et al.~\cite{kwon2020convolutional} adapted CNNs to predict laser power from melt pool images.

However, the introduction of process models (either physics-based or data-driven) introduces several sources of uncertainty such as model parameter uncertainty, input uncertainty, model error, etc., which is then propagated to the QoI predicted by the model~\cite{nath2019uncertainty,berkcan2020Optimization,kapusuzoglu2021information}. Data-driven models are created with data collected from experiments. The uncertainty in the measurement process from instruments, human error, etc. needs to be considered. It is also important to select the optimal model and tuning parameters of the model (e.g., number of layers and units in a deep neural network), and avoid data overfitting. Often the amount of data available to construct the data-driven model is limited, leading to uncertainty in the model prediction. Thus, it is necessary to consider the model uncertainty for an accurate and reliable prediction model. The Bayesian approach has been extensively used to characterize the epistemic uncertainty in the model. Several studies with respect to physics-based AM models have employed Bayesian strategies for model parameter estimation~\cite{nath2020optimization,mahmoudi2018multivariate,berkcan2020Optimization}. However, when the number of parameters to be calibrated is large, use of traditional Bayesian inference for model parameter estimation becomes computationally unaffordable. 
 
Most of the data-driven prediction models in the AM literature have been used for analyzing and understanding the AM process and the relationship between the process parameters and the QoI. Some studies such as Ding et al.~\cite{ding2016bead} used an ANN model to optimize the WAAM process parameters for arriving at the desired bead geometry. However, in AM, several quality characteristics may need to be maximized for a part to be acceptable. For example, a part may need to be geometrically accurate and at the same time have high mechanical strength. Sometimes these objectives are conflicting in nature, e.g., attempting to minimize voids in the part may result in part geometry distortion. Also, since process variability is a major concern and there is model uncertainty in predicted the QoI, it is necessary to optimize the process parameters for \textit{multiple} objectives while also considering these uncertainty sources. 

Based on the above discussion, a data-driven approach for multi-objective optimization under uncertainty of FFF process parameters is developed in this paper, by specifically considering model uncertainty and input variability. The main contributions of this work are as follows:~(1) Experiments are conducted in the laboratory to print parts with different process parameter values, and data is collected to measure two quality characteristics of the parts.~(2) Data-driven Bayesian neural network (BNN) prediction models (one for each quality characteristic) are constructed by implementing Monte Carlo (MC) dropout in the neural network models to quantify the uncertainty in the model prediction as well as the input variability due to the intrinsic randomness in the process parameters.~(3) The trained BNN models are then used to find the optimal values of process parameters, and several multi-objective problem formulations are investigated for robust design optimization (RDO).

The remainder of the paper is organized as follows. The proposed methodology is presented in Section~\ref{sec:method}, followed by the implementation of the methodology for an FFF process in Section~\ref{sec:results}. Section~\ref{sec:conclusion} provides the concluding remarks.

\section{Methodology}\label{sec:method}

\begin{figure}[ht!]
\centering
\includegraphics[width=0.55\textwidth]{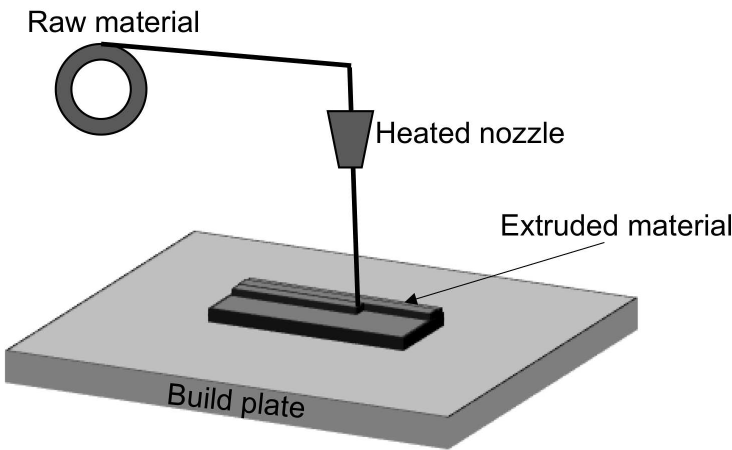}
\caption{Schematic of FFF}
\label{fig:FFFschematic}
\end{figure}

FFF (also known as fused deposition modeling or FDM\textsuperscript{\textregistered}) is an AM technology based on material extrusion~\cite{international2015additive} generally used for manufacturing polymer and plastic parts. A spool of the material in filament form is attached to the AM machine or 3D printer. The continuous strand of material is fed to the heated nozzle which then extrudes the material in a predefined path to form the part in the desired shape. The schematic of the FFF process is shown in Fig.~\ref{fig:FFFschematic}.
In this section, we present the methodology for process parameter optimization under uncertainty with a focus on the FFF process. However, the proposed methodology is applicable to any AM process with corresponding data and QoI prediction models. The three components of the methodology are:~(a) Data collection,~(b) Construction of data-driven prediction models including uncertainty, and~(c) Multi-objective optimization under uncertainty.

\subsection{Data collection}\label{sec:datacollection}

For a data-driven methodology, the first step is to collect data to build the prediction models for the QoI. Traditionally, data is collected from physical experiments. With the advent of physics-based modeling, some researchers also use data generated by the physics-based model~\cite{wang2019data}. In that case, the physics-based model prediction should be first validated with experiments to ensure that the training data closely approximates the actual physics of the AM process. In some cases, data available from previous studies or in the public domain might also be used~\cite{levine2020outcomes,lane2020measurements,cole2020amb2018,kapusuzoglu2021information}. For this work, we collected data from laboratory experiments to build the prediction models. The shape of the part is conceptualized, and a CAD model is first built and then sliced in a slicing software where the printing path is also defined. The printing instructions thus generated are used to print the part.

Various sensors can be used to monitor the AM process and parts. The most common measurement QoIs are melt pool temperature, part dimension, surface roughness, microstructure, tensile properties, etc. Monitoring can be broadly divided into online and offline monitoring techniques. Online monitoring refers to~\textit{in situ} monitoring of the part during the manufacturing process in order to implement process control. Effective online monitoring is not disruptive to the ongoing process, and is non-destructive to the part such as monitoring of the temperature profile using an infrared (IR) camera, or monitoring of the part geometry using profilometer or optical camera. Offline or~\textit{ex situ} monitoring of the part after it has been produced may be either destructive (such as microstructure characterization with scanning electron microscope) or non-destructive (such as part dimension measurement using calipers). Both contact and non-contact techniques may be used for monitoring. For example, temperature data can be collected through contact thermocouples or non-contact IR cameras. Thus, depending on the QoI and the monitoring mode, the instruments and experimental setup are determined. The collected data is then used to construct the prediction models. In this paper, the QoIs are part thickness and bond quality, both of which are measured offline after the part is manufactured.

\subsection{Construction of data-driven prediction model}\label{sec:datadrivenmodel}

There are several machine learning techniques to build prediction models using data. Commonly used approaches include Decision Tree, K-Nearest Neighbor, Support Vector Machine, Linear Discriminant Analysis, Quadratic Discriminant Analysis, Artifical Neural Network (ANN), Polynomial Chaos Expansion, Radial Basis Function, Gaussian Process Modeling, Random Forest Regression etc. As the amount of data available has increased, use of ANN has gained popularity; several types of ANN are available, such as deep neural network (DNN), recurrent neural network (RNN) to handle time series data, and convolution neural network (CNN) to handle image data. Machine learning (ML) models appear promising for complex systems that are not fully understood or cannot be represented with simplified physics-based relationships, given adequate quality and quantity of data. Generally, the construction of data-driven ML models does not require in-depth knowledge of the complex physics inherent in the physical process~\cite{berkcan2020PIML}. In this work, adequate experimental data is available to build a deep learning (DL) model based on the observation data, thus we pursue the ML approach. In addition, we quantify the DL model uncertainty by constructing Bayesian neural network (BNN) models to predict the QoIs and use these BNN models to perform optimization under uncertainty. In the next sections, we introduce the underlying mechanisms in the feedforward neural network and Bayesian neural network (BNN).

\subsubsection{Neural networks}\label{sec:neuralnet}

\begin{figure}[!htb]
	\centering
	\centerline{\includegraphics[width=0.35\textwidth]{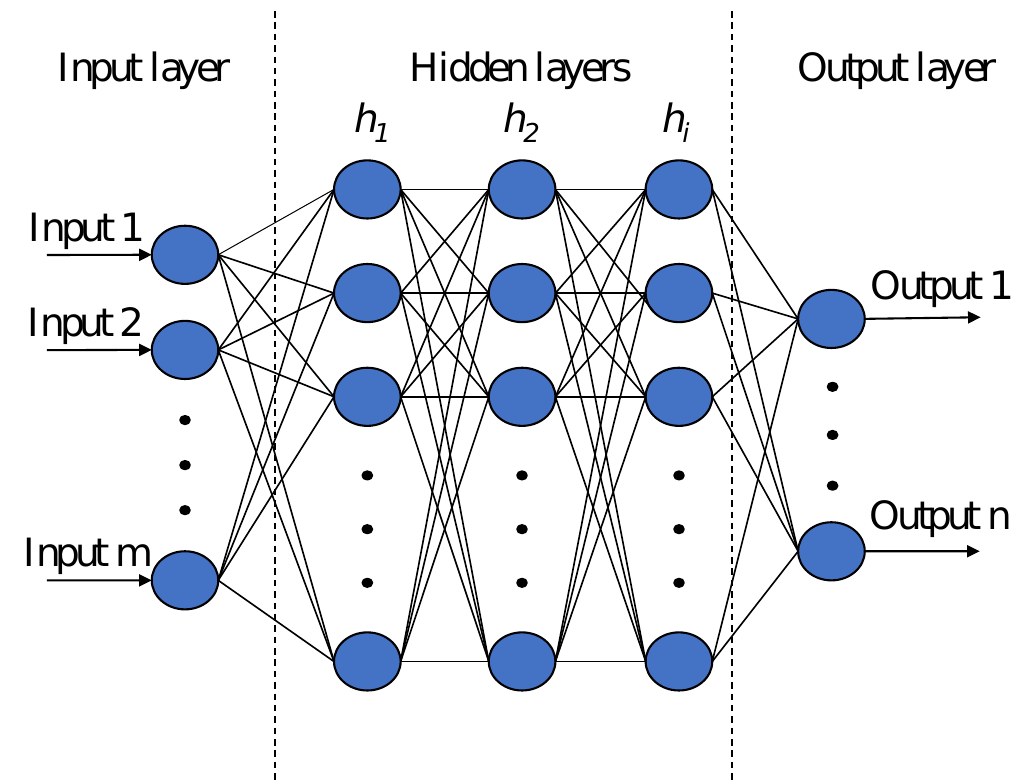}}
	\captionsetup{width=0.90\textwidth}
	\caption{A feedforward neural network with $n$ hidden layers} 
	\label{fig:FeedforwardNetwork}
\end{figure}

Feedforward neural network is the simplest type of ANN. It has four major components: neuron, activation function, cost function, and optimization. The connections between the units of the feedforward neural networks do not form a cycle. The information moves only forward from the input layer through the hidden layers and to the output layer as shown in Fig.~\ref{fig:FeedforwardNetwork}. The values of the various input variables of a particular neuron are multiplied by their associated weights, then the sum of the products of the neuron weights and the inputs are calculated at each neuron. The summed value is passed through an activation function that maps the summed value into a fixed range before passing these signals on to the next layer of neurons.

The predictions of the feedforward neural network after a forward propagation are compared against the actual values by defining a loss function, which measures how far off the predictions are form the observations for the training samples. After the forward propagation, backpropagation algorithms are performed to keep track of small perturbations to the weights that affect the error at the output and to distribute this error back through the network layers by computing gradients for each layer using the chain rule. In order to minimize the value of the loss function, necessary adjustments are applied at each iteration to the neuron weights in each layer of the network. These procedures are performed at each iteration until the loss value converges to a stable value.

ANNs are able to learn powerful representations that can map high dimensional data to an array of outputs. However, deterministic neural networks do not allow for estimating the uncertainty in their predictions. Model prediction uncertainty is important to account for in AM. In addition, there is process variability, uncertainty in measurement, and also limited data due to the high cost associated with conducting experiments especially for metal AM. The quantification of model uncertainty using the Bayesian method is discussed next.

\subsubsection{Bayesian neural network}

In the Bayesian context, the distributions of the model parameters (neuron weights $\mathbf{w}$) that are most likely to generate the observed data are inferred. A prior distribution over the neuron weights $p(\mathbf{w})$ is defined, and a likelihood function $p(\mathbf{Y} | \mathbf{X}, \mathbf{w})$ is defined to represent the probability of generating the observed data given model parameters. Bayesian neural network (BNN) implementations so far have used Gaussian prior distributions $p(\mathbf{w}=\mathcal{N}(0, \mathcal{I}))$ to replace the deterministic network's weight parameters~\cite{denker1991transforming,mackay1992practical,neal2012bayesian}, thus representing the epistemic uncertainty in the DNN model. Following Bayes' theorem, a posterior distribution over the model parameters given the training set $\{\mathbf{X, Y}\} = \{\{\mathbf{x}_1, ...,\mathbf{x}_N\}, \{\{\mathbf{y}_1, ...,\mathbf{y}_N\}\}$ is defined by

\begin{equation}
    p(\mathbf{w} | \mathbf{X}, \mathbf{Y}) = \frac{p(\mathbf{Y} | \mathbf{X}, \mathbf{w}) p( \mathbf{w})}{p(\mathbf{Y} | \mathbf{X})}.
\end{equation}

After inferring the posterior distributions of  $p(\mathbf{w})$, the predictive distribution of the model output for a new input $\mathbf{x^{*}}$ is
\begin{equation}
    p(\mathbf{y^{*}} | \mathbf{x^{*}}, \mathbf{X}, \mathbf{Y}) = \int_{\Omega} p(\mathbf{y^{*}} | \mathbf{x^{*}}, \mathbf{w}) p(\mathbf{w} | \mathbf{X}, \mathbf{Y}) d\mathbf{w}.
\end{equation}
The posterior distribution of model parameters $p(\mathbf{w} | \mathbf{X}, \mathbf{Y})$ cannot be evaluated analytically over the whole parameter space $\Omega$ due to the highly non-linear behavior in the neural network caused by the non-linear activation functions and their combinations across multiple hidden layers. Thus, it becomes difficult to perform exact analytical inference in BNNs. 

Several approximate inference techniques are available to infer the posterior distribution $p(\mathbf{w} | \mathbf{X}, \mathbf{Y})$, such as variational inference~\cite{blundell2015weight}, Approximate Bayesian Computation~\cite{beaumont2002approximate}, Markov Chain Monte Carlo (MCMC)~\cite{gilks2005m}, and Particle Filter~\cite{arulampalam2002tutorial} methods. Markov Chain Monte Carlo (MCMC) can be a more accurate technique than the other methods, but it is computationally too expensive. It would not be affordable to perform MCMC to estimate posterior distributions of the deep neural network model parameters (i.e., weights) since there are usually a large number of parameters in the model. Variational inference (VI) fits a simple and tractable distribution $q_{\theta}(\mathbf{w})$ to the posterior, parametrized by a variational parameter $\theta$~\cite{blundell2015weight}. This approximates the intractable problem by optimizing the parameters of $q_{\theta}(\mathbf{w})$. The accuracy of the variational distribution is often measured by the Kullback-Leibler (KL) divergence between the approximate distribution $q_{\theta}(\mathbf{w})$ and the true model posterior $p(\mathbf{w} |  \mathbf{X}, \mathbf{Y})$. 

Gal and Ghahramani~\cite{gal2016dropout} showed that Monte Carlo (MC) dropout is equivalent to performing approximate VI; the former infers the posterior by performing dropout not only while training a model but also during prediction. In MC dropout, randomly chosen neurons are temporarily removed from the network along with their connections. Next, the gradients of neuron weights are calculated on each smaller neural network and these gradients are then averaged over the training sets to obtain the weights of overall network. The construction of BNNs with MC dropout builds on the concept of dropout as regularization on neural networks. However, in contrast to standard neural networks, MC dropout performs dropout and generates random samples following a Bernoulli distribution for each neuron in the input and hidden layers during prediction. The dropout is applied to the neuron that takes the value $0$ with a given dropout probability $p_d$. The outputs of the network are predicted using the collection of generated random samples from the posterior predictive distribution and the uncertainty in the prediction is quantified. The computational effort for the construction of BNNs with MC dropout is comparable to the standard neural networks. Moreover, the simplicity of the MC dropout strategy provides an efficient way of Bayesian inference to quantify the model prediction uncertainty with a variety of neural networks, such as DNN, CNN, and RNN.

\subsubsection{Uncertainty quantification in Bayesian neural network}

Various sources of uncertainty can be considered, such as (a) Epistemic uncertainty due to lack of knowledge, and (b) Aleatory uncertainty due to the inherent variability across multiple samples and over space and time. Epistemic uncertainty is caused by insufficient knowledge or information about the model and data. In the case of a DNN model, the values of the model parameters such as neuron weights are estimated from the data. If limited data is available, there is epistemic uncertainty in the model prediction. BNN is used to describe the epistemic uncertainty caused by the model by placing distributions over the network weights.

Aleatory uncertainty is caused by the natural variability in the AM process, leading to variability in the process output QoI. For example, the AM parts printed with the same process parameters may have different geometric dimensions and mechanical properties. The observation noise parameter $\sigma$ also needs to be tuned. Similar to the procedure in Bayesian model calibration, observation noise can be learned through the BNN model by minimizing the following loss function~\cite{kendall2017uncertainties}:
\begin{equation}\label{eq:gausslog}
    \mathcal{L}_{BNN}(\theta) = \frac{1}{N}\sum_i \frac{1}{2} \hat{\sigma}^{-2}_i ||\mathbf{y}_i - \hat{\mathbf{y}}_i ||^2 + \frac{1}{2}\text{log}\ \hat{\sigma}^{-2}_i + \lambda g(\mathbf{w})
\end{equation}
where $N$ is the number of observations $\mathbf{y}_i$; $\hat{\mathbf{y}}_i$ and $\hat{\sigma}^{2}_i$ are the predictive mean and observation noise corresponding to input indexed by $i$, the first term represents the norm of residuals for measuring goodness of fit; $\lambda$ is the weight decay parameter, and $g(\mathbf{w})$ represents a regularization function for the weights (or the penalty term which often uses $L_2$ regularization). The second regularization term prevents the network from predicting infinite uncertainty, thus zero loss. The neuron weights can be drawn from the approximate posterior $\hat{\mathbf{w}}\sim q(\mathbf{w})$ to obtain the model outputs of a BNN denoted as $\mathbf{f}^{\hat{w}}(\mathbf{x}) = [\hat{\mathbf{y}}, \hat{\sigma}^{2}]$. The second term of the loss function can be regarded as an uncertainty regularization term and it prevents the network from predicting infinite uncertainty. In order to have a numerically stable network during training, the log variance $\text{log}\ \hat{\sigma}^{2}_i$ is predicted instead of predicting $\hat{\sigma}^{2}_i$. (Sometimes the process parameter settings specified by the designer (such as the nozzle temperature, nozzle speed, layer thickness, etc.) may not be actually realized in manufacturing (this is input uncertainty (epistemic), i.e., the input value specified in the model is different from what is actually in the manufacturing process, and we do not know what the actual value is).  However, this type of uncertainty is not considered in this paper).

The predictive mean and variance are estimated by collecting the results of stochastic forward passes through the model. The mean prediction of the model with $T$ MC samples can be approximated by

\begin{equation}\label{eq:BNNpredmean}
    \mathbb{E}(\mathbf{y}) \approx \frac{1}{T} \sum_{t=1}^{T} \hat{\mathbf{y}}_t,
\end{equation}
and the variance of the prediction is estimated by

\begin{equation}\label{eq:BNNpredvar}
    \text{Var}(\mathbf{y}) \approx \underbrace{ \frac{1}{T} \sum_{t=1}^{T} \hat{\mathbf{y}}_t^2 - \left(\frac{1}{T} \sum_{t=1}^{T} \hat{\mathbf{y}}_t \right)^2}_{\parbox{3cm}{\centering model uncertainty\\[-4pt] (epistemic)}} + \underbrace{\frac{1}{T} \sum_{t=1}^{T} \hat{\sigma}_t^2}_{\parbox{1.5cm}{\centering observation\\[-4pt] noise\\[-4pt] (aleatory)}}.
\end{equation}
In the next section we discuss how to use the model predictions given by Eq.~\ref{eq:BNNpredmean} and Eq.~\ref{eq:BNNpredvar} for optimization under uncertainty.

\subsection{Multi-objective optimization under uncertainty}\label{sec:optimization}

As mentioned earlier, often in AM it is necessary to optimize the process parameters for multiple objective functions. A generic formulation of deterministic multi-objective optimization for $n_{obj}$ objectives may be written as:

\begin{equation}
    \begin{aligned}
    & \underset{\mathbf{x}}{\text{minimize}}
    & & \big\{f_1(\mathbf{x, p}), ..., f_{n_{obj}}(\mathbf{x, p})\big\} \\
    & \text{subject to}
    & & g_i(\mathbf{x, p})\leq 0,\ i = 1, 2,...,n_{con} \\
    & 
    & & \mathbf{lb}_{\mathbf{x}} \leq \mathbf{\mathbf{x}} \leq \mathbf{ub}_{\mathbf{x}} \\
    \end{aligned}
\label{eq:MultiObjFormulation}
\end{equation}
where $\mathbf{x}$ and $\mathbf{p}$ are the design and non-design variables, $g_i,\ i = 1, 2,...,n_{con}$ represent the $n_{con}$ deterministic constraints, and $\mathbf{lb}_{\mathbf{x}}$ and $\mathbf{ub}_{\mathbf{x}}$ are the lower and upper bounds for the design variables $\mathbf{x}$.

\subsubsection{Formulation of multi-objective optimization}\label{sec:optimizationform}

The objective functions in Eq.~\ref{eq:MultiObjFormulation} depend on the intended use of the additively manufactured part. Since optimization involves evaluation of the objective function repeatedly at different process parameter settings, instead of conducting expensive physical experiments the objective functions are evaluated using the data-driven model discussed in Section~\ref{sec:datadrivenmodel}. The prediction from the BNN model has a mean given by Eq.~\ref{eq:BNNpredmean} and the corresponding variance given by Eq.~\ref{eq:BNNpredvar}. The conventional deterministic optimization does not consider the uncertainty in the input variables, data and model parameters; the output of the manufacturing process is sensitive to these uncertainty sources. Optimization under uncertainty can be pursued in two directions:~(1) robust design optimization (RDO)~\cite{zaman2011robustness}, and~(2) reliability-based design optimization (RBDO)~\cite{du2004sequential}. In RDO, both the mean and the variability of the objective function are optimized (since minimizing the variability makes the objective insensitive to variations of the input variables and parameters), and the constraints are satisfied within specified uncertainty bounds. On the other hand, in RBDO, a desired target level of reliability is maximized (i.e., probability of satisfying a desired threshold of performance or quality) by optimizing the decision variables, or a cost function is minimized while satisfying a reliability constraint. In this work we are interested in a robust design to improve the quality of products and processes by optimizing both the mean and variance of the quantities of interest. Therefore, robust design optimization (RDO) is used in this study for the design of the FFF process parameters. (Note also that for well-designed practical systems, the probability of failure would be very low; thus the RBDO formulation would require substantially more function evaluations in comparison to the RDO formulation where the means and variances of the objective function and the constraints can be evaluated with a much smaller number of function evaluations). The robustness of the objective function can be achieved by simultaneously optimizing the mean and variance. Thus RDO even with respect to a single objective QoI becomes a bi-objective optimization problem with two objectives: (a) Optimize the mean of the QoI, and (b) Minimize the variance of the QoI. The resulting bi-objective RDO problem can be approximately solved through a single objective formulation, using a weighted sum approach, i.e., mean and variance terms have weights that reflect the designer’s preference.

Consider the case where the objective is to minimize a function $f(\mathbf{x, p})$ with design variables $\mathbf{x} = [\mathbf{x}_d, \mathbf{x}_{\theta}] $ and non-design variables $\mathbf{p} = [\mathbf{p}_d, \mathbf{p}_{\theta}]$, where $\mathbf{x}_d, \mathbf{p}_d$ are deterministic variables, and $\mathbf{x}_\theta, \mathbf{p}_\theta$ are stochastic design and non-design variables respectively. Since the prediction from the BNN model is stochastic, the single objective RDO formulation using weighted sum approach can be written as
\begin{equation}
    \begin{aligned}
    & \underset{\mathbf{x}  \in \mathbb{R}^{n_x}}{\text{minimize}}
    & & w_1\ \mu_f(\mathbf{x,p}) + w_2\ \sigma_f(\mathbf{x,p}); \\
    & \text{subject to}
    & & \mathbf{lb}_{\mathbf{g}} + \mathbf{k_{c}}\mathbf{\sigma_{g}}{\mathbf{g}(\mathbf{x,p})} \le {\mathbf{g}(\mathbf{x,p})} \le \mathbf{ub}_{\mathbf{g}} - \mathbf{k_{c}}\mathbf{\sigma_{g}}{\mathbf{g}(\mathbf{x,p})},\ \\
    & 
    & & \mathbf{lb}_{\mathbf{x}} \leq \mathbf{\mathbf{x}} \leq \mathbf{ub}_{\mathbf{x}} \\
    \end{aligned}
\label{eq:bi-obj}
\end{equation}
where $\mu_f(\cdot)$ and $\sigma_f(\cdot)$ are the mean and standard deviation of $f$, $0 \leq w_1, w_2 \leq 1$ are the weights representing the relative importance of each objective function, $\mathbb{R}^{n_x}$ represents the design space, $\mathbf{lb}_{\mathbf{x}}, \mathbf{ub}_{\mathbf{x}}$ represent the lower and upper bounds for the design variables, $\mathbf{g}(\mathbf{x,p})$ is the vector of inequality constraints, and $\mathbf{lb}_{\mathbf{g}}, \mathbf{ub}_{\mathbf{g}}$ represent the lower and upper bounds of the constraints. For stochastic constraints, the feasible region is reduced by $k_{c}\sigma_{g}$ in each direction where $\mathbf{k_{c}}$ is a vector of user-defined constants based on the design requirements and $\mathbf{\sigma_{g}}$ is the vector of standard deviations of the constraints~\cite{zaman2011robustness}. For a deterministic constraint, $k_{c}\sigma_g = 0$. Note that in this paper, the optimization needs to consider multiple objectives, considering multiple QoIs. The same approach in the above equation can be extended for $n_{obj}$ objectives by assigning corresponding weights $w_1, w_2, \cdots, w_{n_{obj}}$ to the objective functions.

The computation of objectives $f_j(\mathbf{x, p})$ ($j = 1,\dots, n_{obj}$) and constraints $\mathbf{g}(\mathbf{x,p})$ is affected by uncertainty sources such as input variability and model uncertainty; thus we have stochastic objectives and constraints. The input variability is already present in the experimental data that is used for training the DL model, and further, the model uncertainty (epistemic) is also captured through the BNN approach. The aleatory uncertainty is learned by minimizing the loss function shown in Eq.~{\ref{eq:gausslog}}. MC dropout, which is performing dropout during prediction, is used to quantify the epistemic uncertainty in the model prediction. As a result, the output of the BNN model is stochastic, which incorporates both sources of uncertainty mentioned above. The optimization formulation in Eq.~\ref{eq:bi-obj} properly accounts for the resulting uncertainty. In the next section, we discuss the solution of the optimization problem.

\subsubsection{Solution of multi-objective optimization}
Depending on the design variables and constraints, the optimization problem discussed in Section~\ref{sec:optimization} can be solved using a suitable global optimization algorithm. In this paper, we choose to employ Non-dominated Sorting Genetic Algorithm II (NSGA-II)~\cite{deb2002fast}.

\noindent \underline{Non-dominated Sorting Genetic Algorithm II}

Since a multi-objective problem is formulated, the solution depends on the weighting coefficients. The trade-off among $n_{obj}$ objective functions $f_k,\ k = 1, 2,...,n_{obj}$ can be represented by the Pareto front. The Pareto front is a set of Pareto optimal solutions, which correspond to the solutions for different values of the weighting coefficients for the optimization problem specified in Eq.~\ref{eq:bi-obj}. The NSGA-II algorithm is one of the frequently applied multi-objective optimization evolutionary algorithms for generating the Pareto front~\cite{deb2002fast}. 

The NSGA-II procedure for finding the Pareto front can be briefly described in following steps:
\begin{enumerate}
    \item An initial random population is generated and the fitnesses of the individuals are evaluated during several generations;
    \item Several Pareto fronts are generated by ranking the population based on the non-dominating sorting criteria (where individuals with the best rank represents the first front, the ones with the second best rank generate the second front and so on);
    \item The crowding distance value is assigned to each front once the sorting is completed;
    \item The individuals are selected using a binary tournament selection with crowded-comparison operator (if two individuals have the same rank, the individual with greater crowding distance is selected to increase the diversity, otherwise an individual with a better rank is chosen);
    \item Binary crossover and polynomial mutation are used to generate a new offspring population combined with the current population;
    \item The next generation is set by selection until the population size exceeds the current population; and
    \item The above steps are repeated until the stopping condition is met and a set of non-dominated Pareto optimal solutions are obtained.
\end{enumerate}

The \emph{Pareto optimal set} is the set of all possible Pareto optimal vectors and there is a need for a stopping criterion that evaluates the quality of the Pareto front solutions. There are methods that apply a stopping criterion, such as NSGA-II~\cite{deb2002fast}. Another important concept is the ideal stopping generation criterion, where the solution is as close to the optimal Pareto front as possible in a short amount of generation. The ideal stopping generation is a compromise between the distance to the Pareto optimal front and the cost of generation. Therefore, it is important to be able to determine how good a solution is compared to the optimal one by using an indicator, such as hypervolume indicator~\cite{zitzler2007hypervolume}, and epsilon indicator~\cite{knowles2006tutorial,zitzler2003performance}.

\noindent \underline{Evaluating performance: hypervolume indicator}\label{sec:performanceindicator}

Among the variety of performance indicators for genetic algorithm in optimization, the \emph{hypervolume indicator} has been favored by many researchers to measure the quality of a solution~\cite{zitzler2007hypervolume,zitzler2000comparison}. The hypervolume indicator can capture the closeness of the solutions to the Pareto optimal set and partially the spread of the solutions across the objective space.

The hypervolume indicator, $\mathcal{I}_{hv}(\mathcal{A})$, computes the volume of the region, $H$, defined by a set of reference or nadir points and given set of points, $\mathcal{N}$ and $\mathcal{A}$, respectively as 
\begin{equation}
    \mathcal{I}_{hv}(\mathcal{A}) = \text{Volume}\Bigg(\bigcup_{\forall\boldsymbol{a}\in\mathcal{A};\forall\boldsymbol{n}\in\mathcal{N}} \text{hypercube}(a, n) \Bigg),
\end{equation}
where larger values of $\mathcal{I}_{hv}(\mathcal{A})$ corresponds to better solutions. The absolute performance of an optimization algorithm is measured using nadir points, which are the worst elements of the Pareto front solutions. 

The performance of an algorithm as the evolution proceeds can be tracked by transforming the indicator~\cite{knowles2006tutorial}:
\begin{equation}
    \mathcal{I}_{hv}(t) = \mathcal{I}_{hv}(\mathcal{P}_t) - \mathcal{I}_{hv}(\mathcal{P}_{t-1}),
\end{equation}
where $\mathcal{P}_{t-1}$ and $\mathcal{P}_t$ are the previous and current non-dominated elements of the local Pareto optimal front.

\subsubsection{Experimental validation}
The subsection discusses two types of validation as part of the methodology: first the prediction model is validated against experimental data, and second, the optimization solution is also validated against experimental data.  (Note that there is also cross-validation during the model construction phase, by partitioning the data into training and testing sets, in order to check that the model has the required accuracy, as explained in Section~\ref{sec:crossval}).

\noindent \underline{BNN model validation}

Quantitative comparison of model prediction against experimental data is traditionally done using hypothesis testing. In recent model validation literature, several quantitative methods have been pursued, such as classical hypothesis testing, Bayesian hypothesis testing, reliability-based method, area-metric-based method, etc. Any of these methods can be used to quantitatively validate the optimization results~\cite{ling2013quantitative}. The classical t-test is used in this paper to test the null hypothesis that the mean of observations is equal to the mean of the model prediction. The $t$-test is based on Student's $t$-distribution and the corresponding test statistic $t$ is
\begin{equation}
    t=\frac{\overline{Y}_D-\mu_m}{s_D/\sqrt{n}}
\end{equation}
where $\overline{Y}_D$ is the sample mean of experimental observations, $\mu_m$ is the mean prediction, and $s_D$ is the sample standard deviation. The $p$-value (i.e., $p=2F_{T,n-1}(-|t|)$, where $F_{T,n-1}$ is the cumulative distribution function (CDF) of a $t$-distribution with (n-1) degrees of freedom) is compared with the significance level $\alpha$ (usually 0.01 or 0.05). If the $p$-value is less than or equal to $\alpha$, then the null hypothesis is rejected, otherwise the null hypothesis is not rejected.

\noindent \underline{Optimization solution validation}

Since we are using an approximate surrogate model, we perform the second validation to check whether the propagation of surrogate model error through the optimization process has significantly affected the optimum solution. Validation of the optimization methodology can be achieved by comparing the performance of the optimum solution against other randomly selected settings for the decision variables. In this paper, the Pareto optimal solutions obtained using the proposed methodology are validated by conducting actual printing experiments at the optimal process parameter settings and other random settings. The performance comparison is done by comparing the quality objectives achieved in the actual manufactured parts, such as mean bond length, mean part thickness, and their standard deviations. Section~\ref{sec:mo_opti} considers four combinations of these individual objectives. In each case, it is investigated whether the optimum solution gives better part quality than randomly selected process parameter settings.

\subsection{Summary of methodology}\label{sec:summary}

\begin{figure*}[!htb]
\centering
\includegraphics[width=0.75\textwidth]{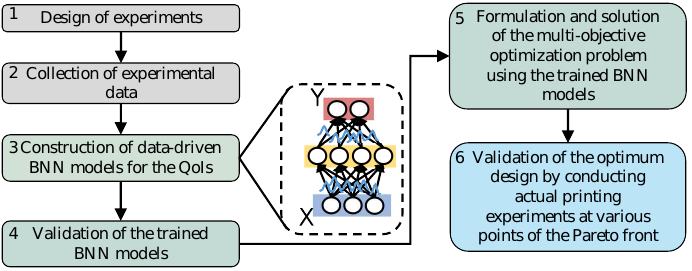}
\caption{Flowchart of the proposed methodology}
\label{fig:summary}
\end{figure*}

The steps of the proposed methodology for multi-objective optimization of AM process parameters under uncertainty are summarized in Fig.~\mbox{\ref{fig:summary}}. The proposed method consists of five main components: (a) Collection of experimental data, (b) Construction of data-driven probabilistic prediction models, (c) Validation of the trained BNN models, (d) Multi-objective optimization under uncertainty of process parameters, and (e) Validation of the optimization results.

The methodology presented in this paper can be generalized for any AM process, as long as experimental data is available to build the data-driven model. In the next section, we demonstrate the effectiveness of the proposed methodology for four different cases with multiple objectives.

\section{3D printing example}\label{sec:results}

In this section we demonstrate the implementation of the proposed methodology on a part of dimensions  $35$ mm $\times$ $12$ mm $\times$ $4.2$ mm printed with ABS (acrylonitrile butadiene styrene) using an Ultimaker S5 printer~\cite{ultimakerS5}. The objective is to find the optimal process parameters nozzle temperature ($T_e$), nozzle speed ($V_e$), and layer height ($l_t$), i.e., ${\boldsymbol{x}} = [{T_e},{V_e},{l_t}]$ such that both the dimensional accuracy and bond quality of the part are maximized. Since the maximum extrusion volume for the 0.8 mm diameter nozzle used in the experiments is 24 mm$^3$/s, an inequality constraint $g = V_e \times l_t \times L_w \le 24$ mm$^3$/s, where $L_w = 0.8$ mm is the raster width, is added to the optimization problem.

\subsection{Data collection from experiments}

\begin{figure}[!htb]
	\centering
	\centerline{\includegraphics[width=0.35\textwidth]{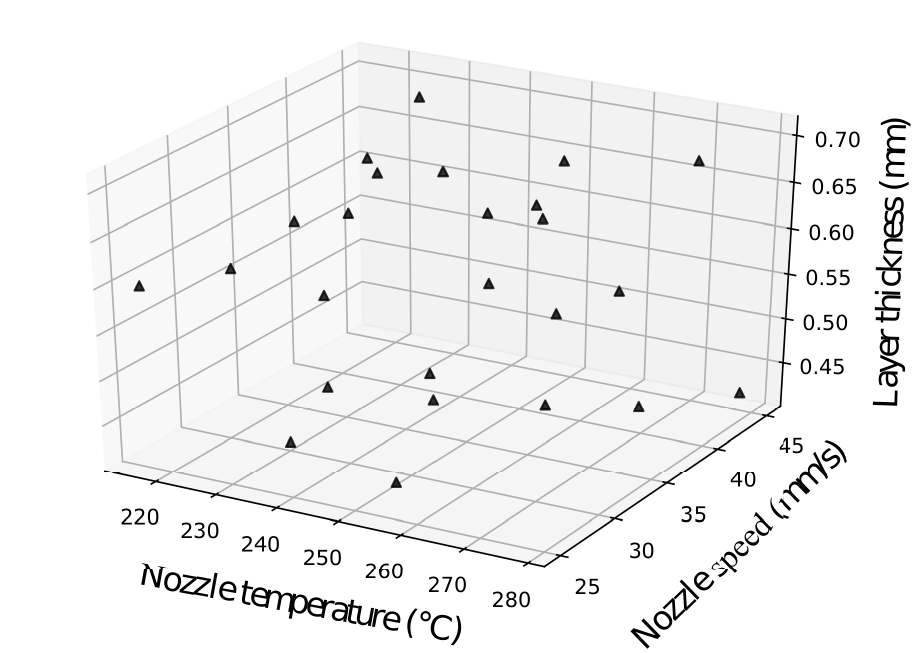}}
	\captionsetup{width=0.90\textwidth}
	\caption{Design of experiments for process parameters} 
	\label{fig:LHS}
\end{figure}

In this work, we use ABS and modify the printing environment by adding an enclosure to the 3D printer to isolate the printer from environmental effects  (see \mbox{\cite{berkcan2020Optimization,berkcan2020PIML,nath2019uncertainty,nath2020optimization}} for details). Using Latin hypercube sampling (Fig.~\ref{fig:LHS}), 25 sets of process parameters $\boldsymbol{x}$ are generated and three parts are printed at each parameter setting. The ranges considered for the variables are: $T_e : 215^{\circ}$C to $280^{\circ}$C, $V_e:$ 25 mm/s to $45$ mm/s, and $l_t=\{0.42,0.60,0.70\}$ mm. 
\begin{figure}[!htb]
	\centering
	\centerline{\includegraphics[width=0.4\textwidth]{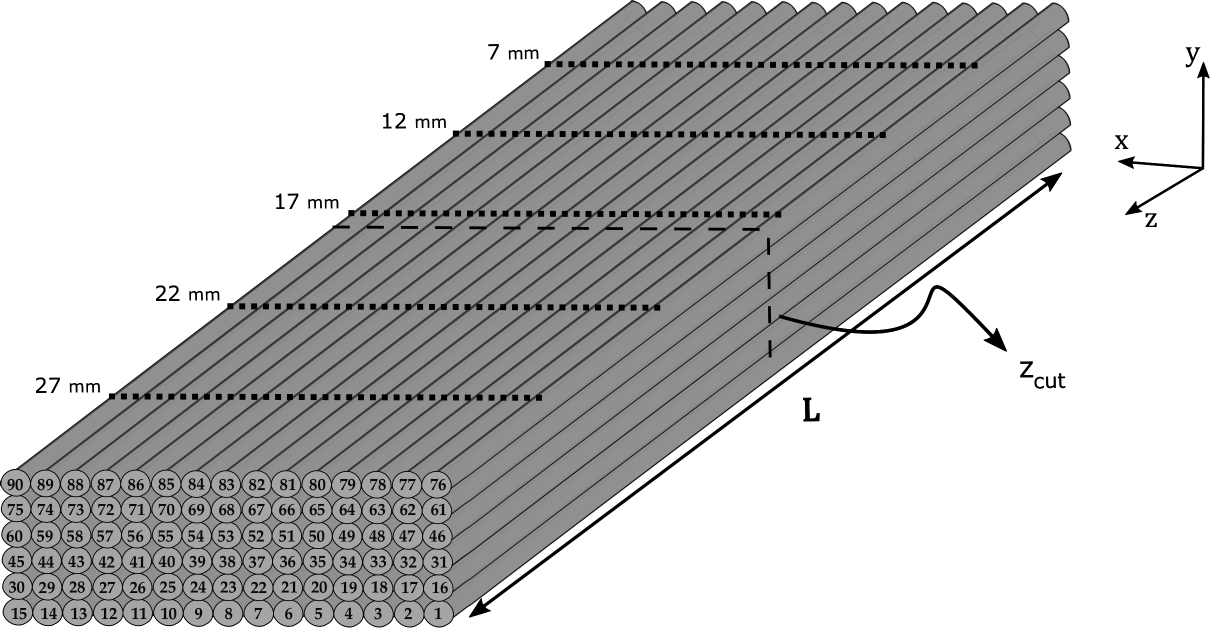}}
	\captionsetup{width=0.90\textwidth}
	\caption{Deposition sequence of unidirectional filaments} 
	\label{fig:DeposSeqSequential}
\end{figure}

\begin{figure}[!htb]
	\centering
	\centerline{\includegraphics[width=0.15\textwidth]{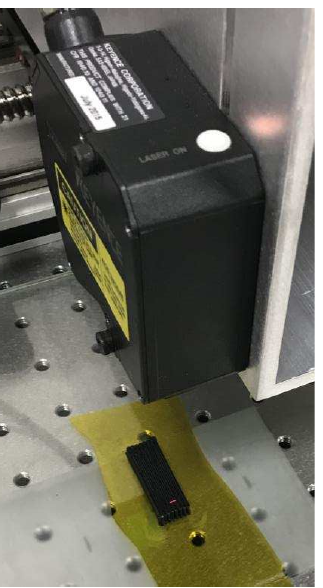}}
	\captionsetup{width=0.90\textwidth}
	\caption{Thickness measurement of the printed part} 
	\label{fig:thickness_measurement}
\end{figure}

Since the focus is on maximizing the dimensional accuracy and bond quality, data pertaining to these two quality characteristics are collected. The measure for dimensional accuracy is considered to be the error in part thickness, i.e., the difference between the printed part thickness and the target part thickness of $4.2$ mm. As shown in Fig.~\ref{fig:thickness_measurement}, using a laser displacement sensor Keyence LK-H057~\cite{LKsensor} the part thickness is measured at $z = \{7, 12, 17, 22, 27\}$ mm at discrete points along the x-axis. The bond lengths (mesostructural feature of interest) between the filaments were measured at cross-section $z_{\rm cut}=L/2=17.5$ mm as shown in Fig.~\ref{fig:DeposSeqSequential} with the use of microscopy images processed through the ImageJ software~\cite{schneider2012nih}.

\begin{figure}[!htb]
\centerline{\includegraphics[width=.45\columnwidth,keepaspectratio]{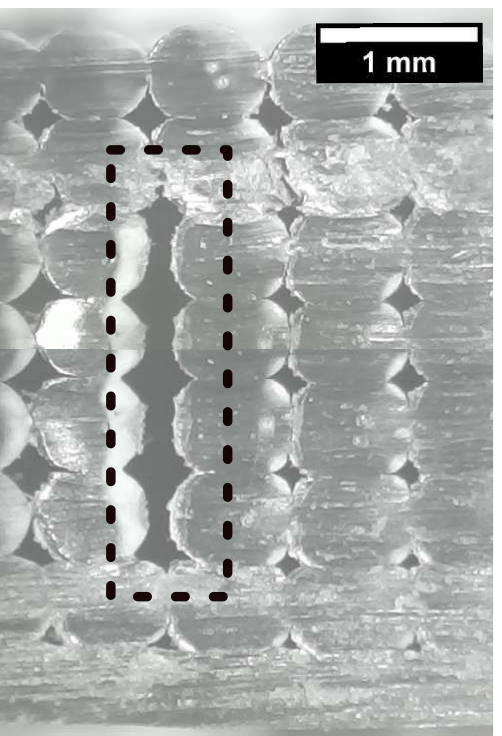}}
	\caption{Cross-section view of a sample with poor bonding} 
	\label{fig:part5delamination}
\end{figure}

The bond length measurements at each interface of a part, which is printed with inputs $(T_e,V_e,l_t)=(227^{\circ}$C, 41\ mm/s, 0.6 mm), are given in Table.~\ref{table:part5_bondlengths}. The interfaces are numbered from left to right for all layers (e.g., the label for the interface between filaments 1 and 2 is 1 and the label for the interface between filaments 89 and 90 is 14 in Fig.~\ref{fig:DeposSeqSequential}). 
\begin{table*}[!htb]
\small
\setlength\tabcolsep{1.2pt}
    \caption{Bond length measurements (in mm) at each layer for $(T_e,V_e,l_t)=(227^{\circ}$C, 41\ mm/s, 0.6 mm)}
    \label{table:part5_bondlengths}
\centering{%
    \begin{tabular*}{\linewidth}{@{\extracolsep{\fill}}l*{14}c@{\extracolsep{\fill}}}
    \toprule
            Layer & \multicolumn{13}{c}{Interface}\\
            &  1 & 2 & 3 & 4 & 5 & 6 & 7 & 8 & 9 & 10 & 11 & 12 & 13 & 14 \\
    \midrule
            1 & 0.383 & 0.381 & 0.416 & 0.469 & 0.459 & 0.487 & 0.472 & 0.490 & 0.528 & 0.525 & 0.500 & 0.474 & 0.452 & 0.421  \\
            2 & 0.431 & 0.444 & 0.452 & 0.429 & 0.396 & 0.431 & 0.322 & 0.360 & 0.294 & 0.353 & 0.363 & 0.408 & 0.317 & 0.342  \\
            3 & 0.462 & 0.495 & 0.434 & 0.487 & 0.431 & 0.345 & \bfseries 0.000 & 0.167 & 0.101 & 0.180 & 0.281 & 0.347 & 0.251 & 0.220 \\
            4 & 0.424 & 0.464 & 0.416 & 0.444 & 0.439 & 0.314 & 0.170 & \bfseries 0.000 & 0.220 & 0.261 & 0.248 & 0.238 & 0.210 & 0.185 \\
            5 & 0.365 & 0.441 & 0.419 & 0.441 & 0.431 & 0.281 & 0.162 & \bfseries 0.000 & 0.246 & 0.266 & 0.243 & 0.177 & 0.233 & 0.200 \\
            6 & 0.375 & 0.398 & 0.467 & 0.510 & 0.396 & 0.347 & 0.259 & \bfseries 0.000 & 0.254 & 0.215 & 0.208 & 0.157 & 0.261 & 0.208 \\
            7 & 0.340 & 0.381 & 0.480 & 0.449 & 0.391 & 0.360 & 0.287 & 0.223 & 0.180 & 0.223 & 0.134 & 0.182 & 0.195 & 0.193  \\
    \bottomrule
    \end{tabular*}
}%
\end{table*}
The bonding quality between adjacent filaments at $z_{\rm cut}=17.5$ mm of the same part is illustrated in Fig.~\ref{fig:part5delamination}. The 8th and 9th filaments at layers 3, 4, 5, and 6 show delamination. This delamination phenomenon was also observed at the exact same interfaces in the other two sets of samples printed with the same process parameter values. The average of these measured bond lengths gives the mean bond length of the part.

\begin{figure}[!htb]
\centerline{\includegraphics[width=.85\columnwidth,keepaspectratio]{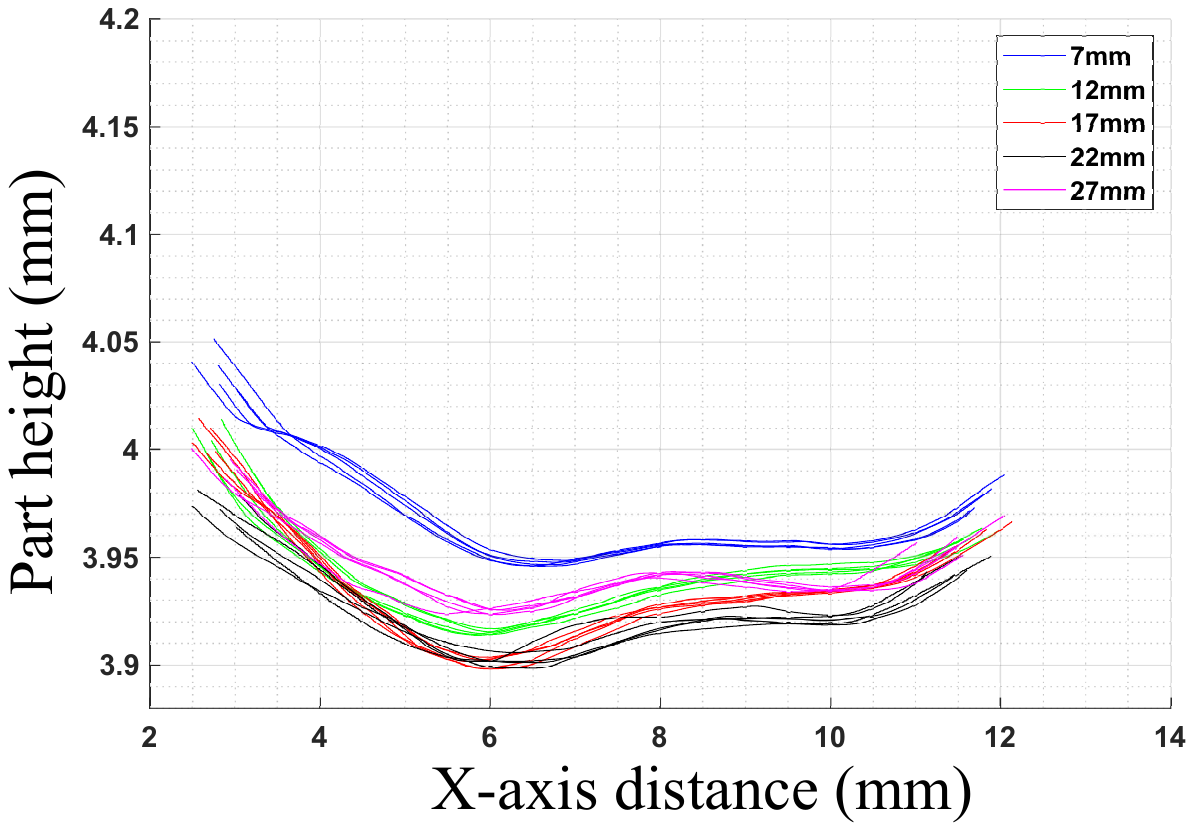}}
	\caption{Part thickness of a sample along Z-axis locations}
	\label{fig:part5thickness}
\end{figure}

The thickness of the part printed at $(T_e,V_e,l_t)=(227^{\circ}$C, 41\ mm/s, 0.6 mm), measured using the laser sensor, is shown in Fig.~\ref{fig:part5thickness}. At each of the five Z-axis points, the measurements are taken along the X-axis to measure the part thickness. It is observed that repeated measurements along X-axis for each Z-axis point show similar values demonstrating the reliability of the measuring system. These discrete point measurements averaged together give the mean thickness of the part.

\subsection{Model training and prediction}
Separate BNNs are constructed for predicting each individual QoI (bond length and thickness in this case). Since the data is collected at many discrete spatial locations, the inputs to the BNNs include the process parameters nozzle temperature, nozzle speed, layer height, and also the spatial locations. The inputs to the bond length model ($\mathrm{BNN}_{\mathrm{bl}}$) are $[{T_e},{V_e},{l_t}, x, y]$ and the inputs to the part thickness model ($\mathrm{BNN}_{\mathrm{th}}$) are $[{T_e},{V_e},{l_t}, x, z]$. The output for $\mathrm{BNN}_{\mathrm{bl}}$ and $\mathrm{BNN}_{\mathrm{th}}$ are bond length and part thickness respectively.

The available experimental data related to the bond length model and part thickness model is divided into 2 sets, namely training set (75\% of all data) and testing set (25\% of all data). The input and output data are normalized prior to the training of the BNN models, and the hyperparameters of these models (e.g., batch size, learning rate, dropout rate etc.) are tuned with grid search. The minimum prediction error during prediction for the bond length model, $\mathrm{BNN}_{\mathrm{bl}}$, is achieved using a dropout rate of 0.1 with 128 batch size, rectified Linear Unit (ReLU) activation function and Adam optimizer with a learning rate of 0.004. The minimum prediction error of $\mathrm{BNN}_{\mathrm{th}}$ is obtained using a dropout rate of 0.1 with 128 batch size, sigmoid activation function and Adam optimizer with a learning rate of 0.005.

\subsection{Model performance and cross validation}\label{sec:crossval}

\begin{figure}[!htb]
\centerline{\includegraphics[width=.75\columnwidth,keepaspectratio]{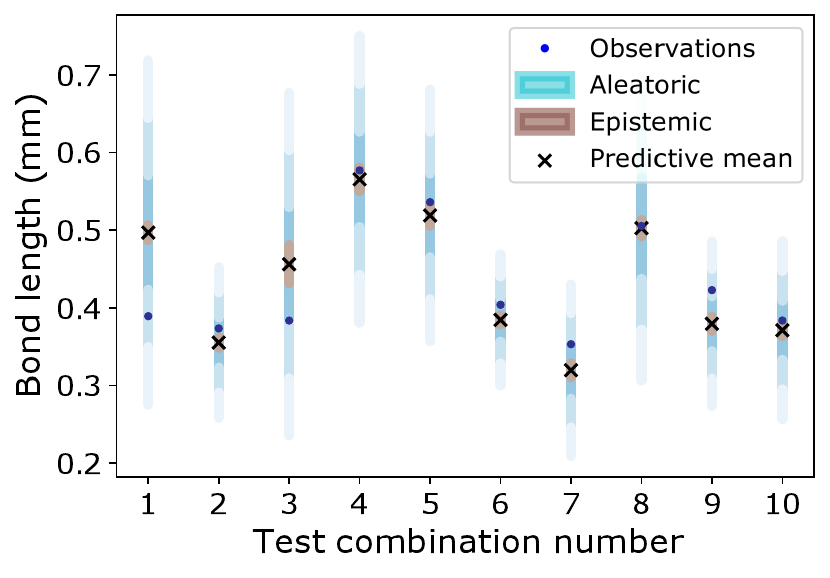}}
	\caption{Predicted bond lengths and uncertainty bounds at randomly selected interfaces from the test data and actual bond length measurements (obtained using microscopy images)}
	\label{fig:temp_vs_bl}
\end{figure}

In order to measure the accuracy of the trained bond length BNN model ($\mathrm{BNN}_{\mathrm{bl}}$), the model predictions for the testing data subset (i.e., subset of 25 percent of all shuffled data set) are compared with the observed bond lengths at randomly selected interfaces of each printed part in Fig.~\ref{fig:temp_vs_bl}. The horizontal axis represents test combinations with different input parameter combinations and the blue dots denote the bond length measurements at the randomly selected interface from the test data for each printed part. In this figure, the black cross denotes the mean prediction of bond length as predicted by the probabilistic model, the shaded light brown area demonstrates the epistemic uncertainty for one standard deviation away from the mean value, the shaded cyan represents the aleatory uncertainty for one standard deviation away from the mean plus the epistemic uncertainty value and the shaded light blue represents the total uncertainty (aleatoric and epistemic) for one standard deviation away from the mean plus the epistemic plus the aleatory uncertainty value. Thus, the whole shaded area represents the uncertainty bounds for three standard deviations away from the mean predictions.

\begin{figure}[!htb]
\centerline{\includegraphics[width=.75\columnwidth,keepaspectratio]{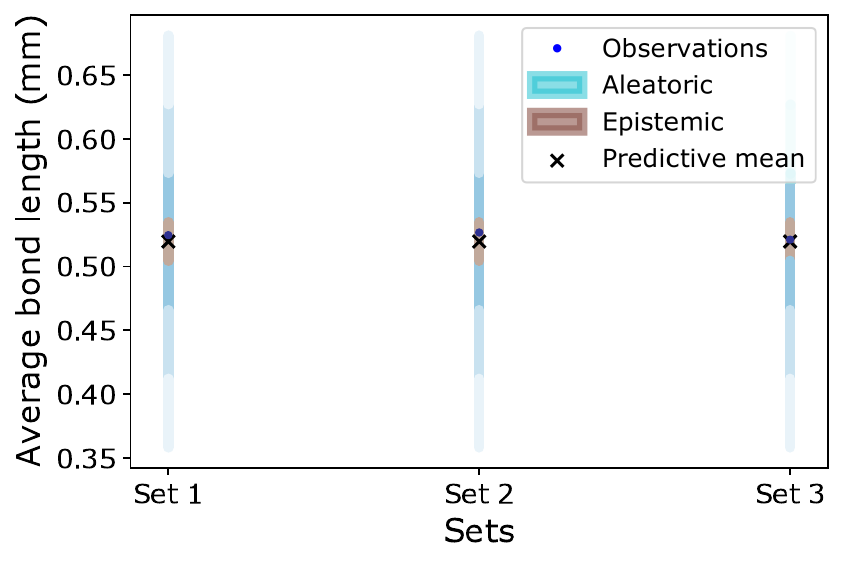}}
	\caption{Average bond length predictions vs. observations of three sets of parts printed with same process parameters} 
	\label{fig:Part25_mean_bl}
\end{figure}

Figure~\ref{fig:Part25_mean_bl} compares the observed average bond lengths with the predictions using $\mathrm{BNN}_{\mathrm{bl}}$. The horizontal axis represents different sets of samples, which are printed using the same process parameters ($T_e,V_e,l_t$). An overall bond quality metric is obtained by averaging the measured and predicted bond lengths of a part at each interface. As shown in Fig.~\ref{fig:Part25_mean_bl}, the model is able to capture the ground truth within half standard deviation.

\begin{figure}[!htb]
\centerline{\includegraphics[width=.7\columnwidth,keepaspectratio]{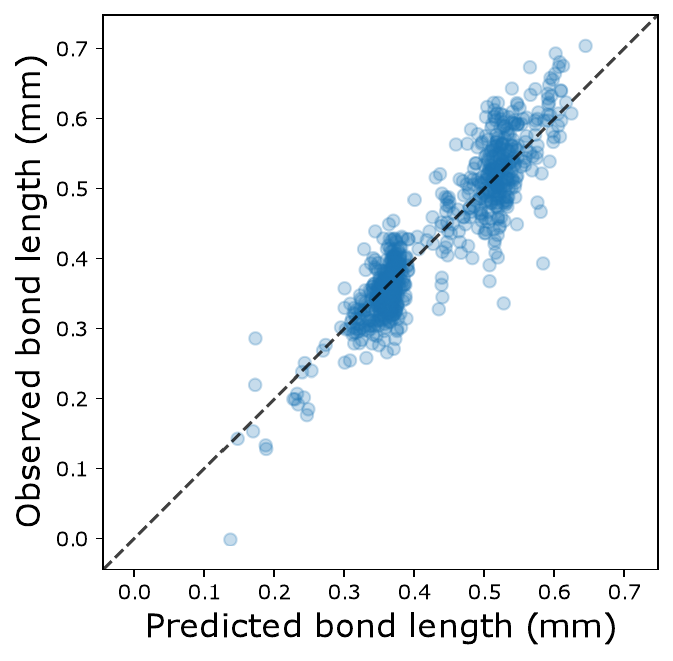}}
	\caption{Comparison of mean dropout prediction and observation of bond lengths} 
	\label{fig:obs_pred_bl}
\end{figure}

\begin{figure}[!htb]
\centerline{\includegraphics[width=.7\columnwidth,keepaspectratio]{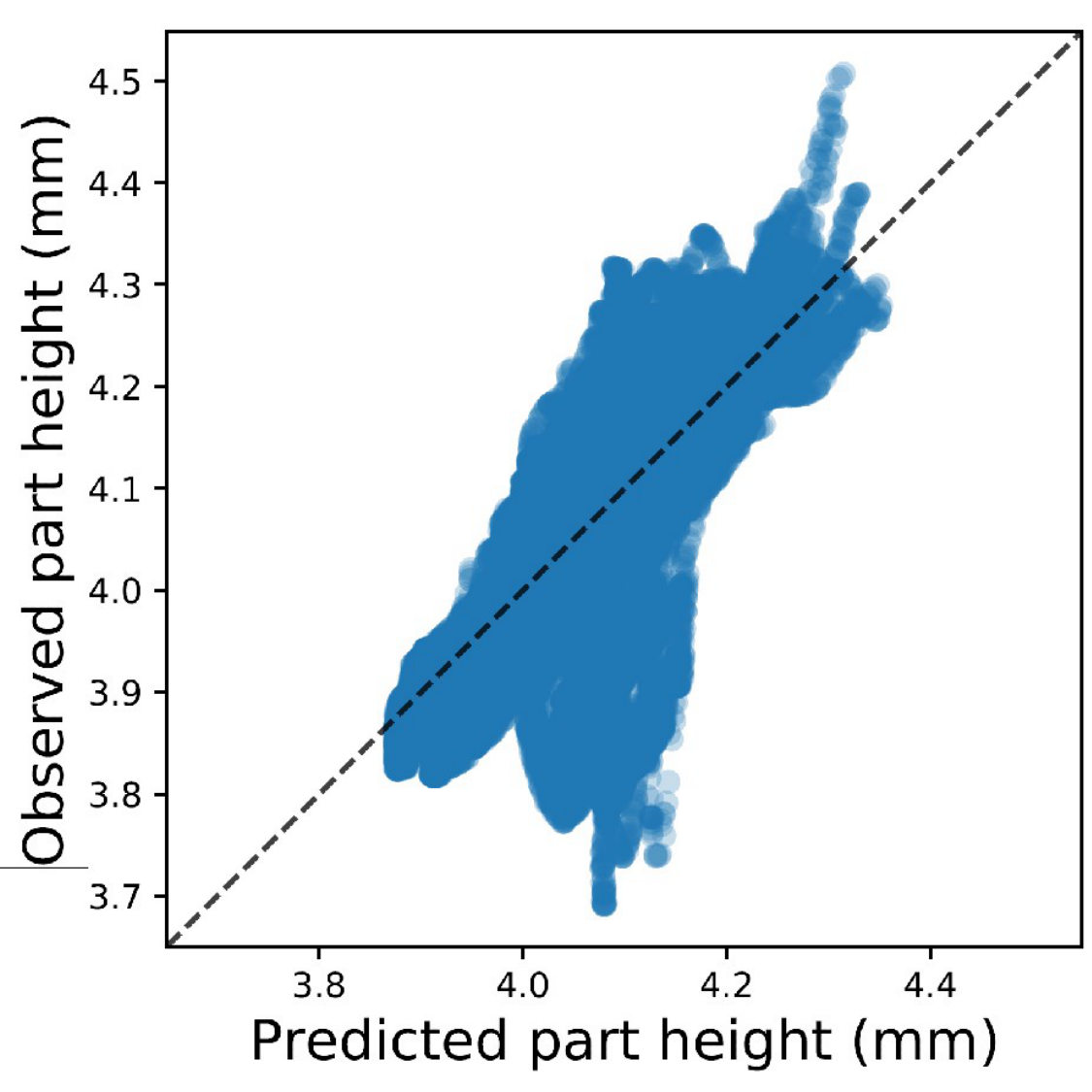}}
	\caption{Comparison of mean dropout prediction and observation of part thickness} 
	\label{fig:obs_pred_th}
\end{figure}

The observed and mean prediction bond length values at the interfaces of the test data using MC dropout are compared on a 45-degree angle line in Fig.~\ref{fig:obs_pred_bl}. The bond length measurements and predictions are carried out at each interface in a layer. The delamination (i.e., zero bond length, which is an outlier) observed in one of the parts is predicted with a non-zero mean bond length. Similar analysis is carried out for part height measurements. Figure~\ref{fig:obs_pred_th} shows the comparison between the observation and mean prediction using MC dropout for the part height. Note that the pairs of the observations and the mean predictions are close to the 45-degree line, showing good agreement between predictions and observations for the $\mathrm{BNN}_{\mathrm{bl}}$ model but not that good for the $\mathrm{BNN}_{\mathrm{th}}$ model. Further cross-validation of the bond length and part thickness prediction models is done using testing data, i.e., (25\% of all shuffled data). The prediction accuracy of both models for the test set is assessed by evaluating the root mean squared error (RMSE); RMSE$=\sqrt{\sum_{i=1}^{N} (y_{i}- \hat{y}_{i})^2/N}$, which is found to be 0.0667 and 0.0826 for $\mathrm{BNN}_{\mathrm{bl}}$ and $\mathrm{BNN}_{\mathrm{th}}$, respectively. RMSE of 0.1 is used as the quantitative criterion for acceptance. The results indicate that the RMSE values for both models are less than 0.1, thus they are accepted for further analysis.

\subsection{Multi-objective optimization}\label{sec:mo_opti}
The predictions from the BNN models are used in the multi-objective optimization methodology to optimize the process parameters such that the dimensional accuracy and bond quality of the part are maximized. Four different optimization cases under uncertainty are considered:\\\\
(1) Case 1, the mean overall bond quality of the part is maximized while minimizing its variance;\\
(2) Case 2, the mean geometric accuracy of the part is maximized while minimizing its variance;\\
(3) Case 3, the mean overall bond quality and the mean geometric accuracy of the part are maximized; and\\
(4) Case 4, the mean overall bond quality and the mean geometric accuracy of the parts are maximized while minimizing the variance in bond quality and geometric accuracy.

Note that cases 1, 2, and 4 include minimization of variance of the part quality (bond quality and geometric accuracy), thus considering uncertainties. Case 3 minimizes the means of two part quality metrics without considering the variance. The weighted sum approach gives a convex combination of two different objectives in the first three cases. Whereas, Case 4 is a convex combination of four different objectives (i.e., one minus the mean value of dimensionless bond length of the part, ($1-\mu_{\mathrm{\hat{bl}}}$), the standard deviation of dimensionless bond length of the part $\sigma_{\mathrm{\hat{bl}}}$, the dimensionless mean and standard deviation of the absolute error between the desired and predicted part thickness ($\mu_{\mathrm{\hat{th}}}$, $\sigma_{\mathrm{\hat{th}}}$)). 

The QoIs (bond length and part thickness) are dependent on the part layer thickness. For example, intra-layer bond length between interfaces is a function of part layer thickness. The maximum achievable intra-layer bond length equals to layer thickness, thus the scale of bond length values changes with changing layer thicknesses. In order to remove this dependency and to prevent the problem caused by different scales, the mean and standard deviation of the QoIs are non-dimensionalized. The mean and standard deviation of the bond length are non-dimensionalized and scaled to unity by dividing with part layer thickness (i.e., nondimensional bond length = (intra-layer bond length)/(layer thickness)). The predicted mean absolute error in part thickness ($\mu_{\mathrm{th}}=|\mu_\mathrm{th_{\mathrm{pred}}}-\mathrm{th}_{\mathrm{desired}}|$, where $\mu_\mathrm{th_{\mathrm{pred}}}$ and $\mathrm{th}_{\mathrm{desired}}$ are the mean dropout prediction of part thickness and desired part thickness, respectively) and standard deviation of the part thickness $\sigma_{\mathrm{th}}$ are also nondimensionalized and scaled to unity as $\mu_{\mathrm{\hat{th}}}=\mu_{\mathrm{th}}/\mathrm{th}_{\mathrm{desired}}$ and $\sigma_{\mathrm{\hat{th}}}=\sigma_{\mathrm{th}}/\mathrm{th}_{\mathrm{desired}}$.

Based on the information obtained from the layer height measurements and the infrared (IR) thermal camera images during the printing process, no significant difference is observed between the measured and reference values; thus, the epistemic uncertainty in the input is not considered. The model uncertainty (epistemic) is described by placing distributions over the model's weights. The aleatory variability (intrinsic randomness) in the input, which can be described as noise in the observations, is present in the experimental data used in training the BNN models and is learned by minimizing the loss function shown in {Eq.~\ref{eq:gausslog}}. As a result, uncertainty in the prediction model as well as the aleatory uncertainty in the input are considered in the optimization.

\begin{figure}[!htb]
\centerline{\includegraphics[width=.85\columnwidth,keepaspectratio]{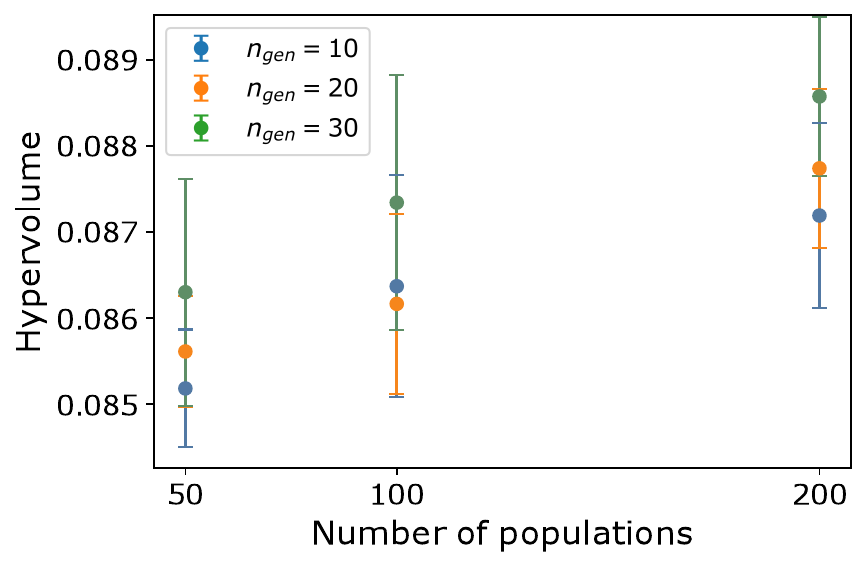}}
	\caption{Hypervolume values (mean and one standard deviation) of nine different experimental designs of Case 3} 
	\label{fig:Hyper_exp_design}
\end{figure}

\begin{figure}[!htb]
\centerline{\includegraphics[width=.85\columnwidth,keepaspectratio]{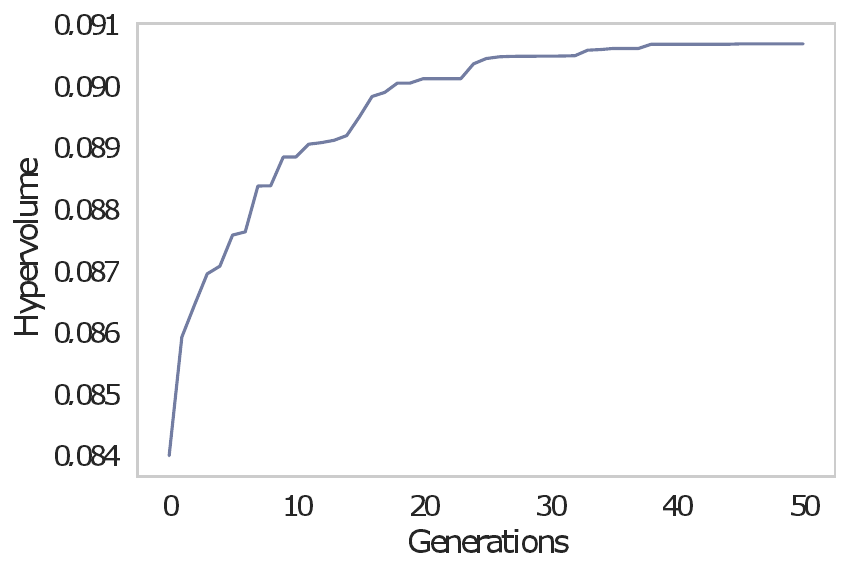}}
	\caption{Hypervolume values vs. number of generations of Case 3 (Population size 200)} 
	\label{fig:hypervolume_vs_gen}
\end{figure}

The hyperparameters of the NSGA-II optimization algorithm (i.e., population size and the maximum number of generations) are tuned by evaluating the performance indicator (hypervolume) (see Section~\ref{sec:performanceindicator}) of nine different experiments. The experiments are repeated 30 times since the methods are stochastic. The convergence of hypervolume values are monitored for each case. The probability of crossover and mutation are chosen as 0.8 and 0.2, respectively. For illustration purposes, the hypervolume values for Case 3 are compared by changing the population size or maximum number of generations while keeping the other one constant (Fig.~\ref{fig:Hyper_exp_design}). The maximum hypervolume value is achieved using a population size of 200 with 30 generations as shown in Fig.~\ref{fig:Hyper_exp_design}. The hypervolume values of other cases show a similar trend. Fig.~\ref{fig:hypervolume_vs_gen} shows in more detail the change in hypervolume values with increasing numbers of generations for a population size of 200 for Case 3. It is found that hypervolume values converge at 30 generations; thus a population size of 200 and 30 generations are used for all the multi-objective optimization cases considered in this paper. The maximum extrusion volume inequality constraint $g$ (Section~\ref{sec:results}) is implemented using a penalty function. The penalty function gives a fitness disadvantage to the individuals that violate the constraint.

\subsubsection{Case 1}
In Case 1, one minus the mean of dimensionless bond quality metric ($1-\mu_{\mathrm{\hat{bl}}}$) and the standard deviation of the dimensionless bond quality metric $\sigma_{\mathrm{\hat{bl}}}$ are minimized.
\begin{equation}
    \begin{aligned}
    & \underset{\mathbf{x}  \in \mathbb{R}^{n_x}}{\text{minimize}}
    & & w_1\ (1-\mu_{\mathrm{\hat{bl}}}(\mathbf{x})) + w_2\ \sigma_{\mathrm{\hat{bl}}}(\mathbf{x}) \\
    & \text{subject to}
    & & {g(\mathbf{x})} = 0.8 mm \times V_e \times l_t \le 24 mm^3/s \\
    & & &215^{\circ}\text{C} \leq T_e \leq 280^{\circ}\text{C} \\
    & & &25 \text{mm/s} \leq V_e \leq 45 \text{mm/s} \\
    & & &l_t \in \{0.42,0.60,0.70\} \text{mm}\\
    \end{aligned}
\label{eq:bi-obj_1}
\end{equation}

\begin{figure}[!htb]
\centerline{\includegraphics[width=.85\columnwidth,keepaspectratio]{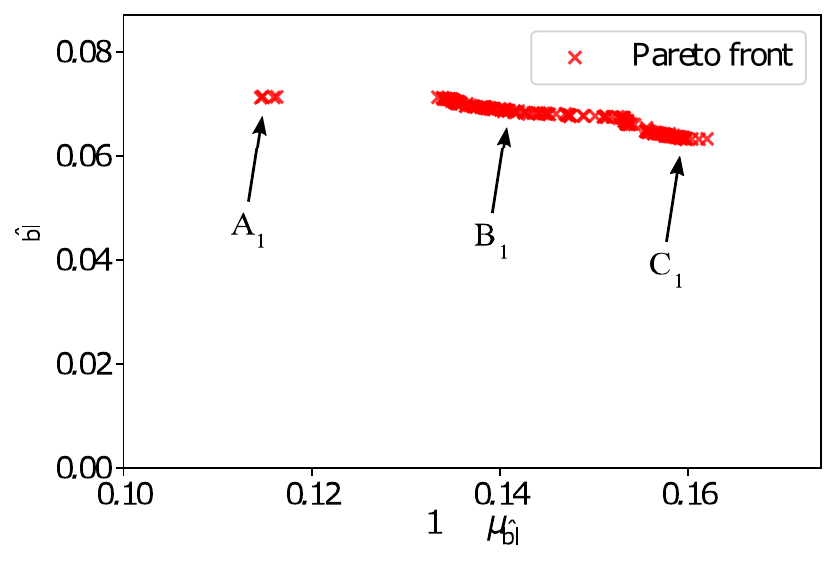}}
	\caption{Pareto front of Case 1: A$_1$ = ($T_e$ ($^{\circ}$C), $V_e$ (mm/s), $l_t$ (mm)) = (217.09, 26.14, 0.42), B$_1$ = (244.54, 29.59, 0.60), C$_1$ = (219.03, 43.96, 0.42)} 
	\label{fig:front_1}
\end{figure}

The Pareto front obtained for these two objectives is demonstrated in Fig.~\ref{fig:front_1} with selected design points A$_1$, B$_1$, and C$_1$ for experimental validation. The design points represent three design variables, i.e., nozzle temperature $T_e$ ($^{\circ}$C), nozzle speed $V_e$ (mm/s), and layer thickness $l_t$ (mm).

In each generation, several values of the design variables (nozzle temperature, nozzle speed, and layer thickness) are generated and passed to the optimizer to evaluate two objectives. The layer height is restricted to three possible values, 0.42, 0.6, and 0.7 mm, in order to have an integer value for the total number of layers given the desired part thickness of 4.2 mm. A single design can be selected from the optimal designs shown in Fig.~\ref{fig:front_1}.

\subsubsection{Case 2}
\begin{figure}[!htb]
\centerline{\includegraphics[width=.85\columnwidth,keepaspectratio]{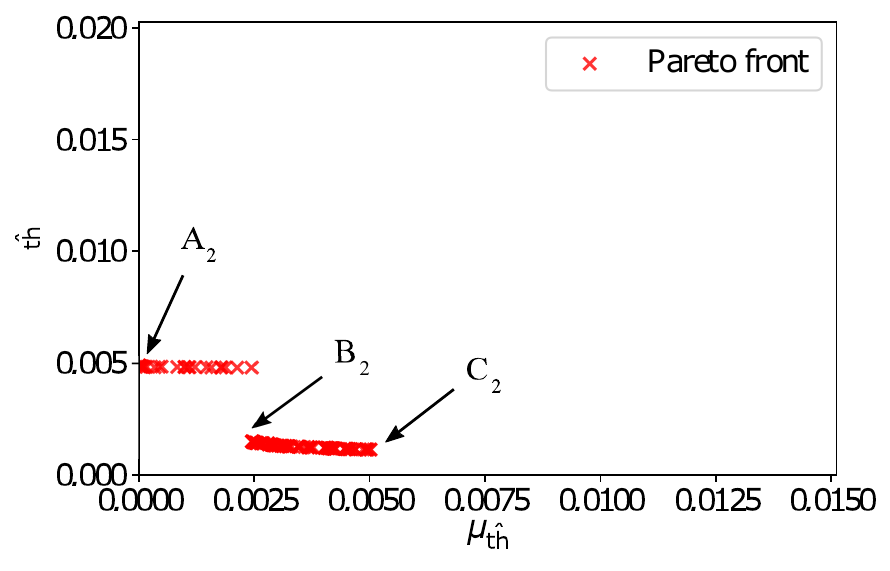}}
	\caption{Pareto front of Case 2: A$_2$ = ($T_e$ ($^{\circ}$C), $V_e$ (mm/s), $l_t$ (mm)) = (272.17, 33.68, 0.42), B$_2$ = (223.92, 31.02, 0.42), C$_2$ = (251.41, 36.63, 0.42)} 
	\label{fig:front_2}
\end{figure}

The dimensionless mean and standard deviation of part thickness error ($\mu_{\mathrm{\hat{th}}}$, $\sigma_{\mathrm{\hat{th}}}$) are minimized in Case 2 using the $\mathrm{BNN}_{\mathrm{th}}$. 
\begin{equation}
    \begin{aligned}
    & \underset{\mathbf{x}  \in \mathbb{R}^{n_x}}{\text{minimize}}
    & & w_1\ \mu_{\mathrm{\hat{th}}}(\mathbf{x}) + w_2\ \sigma_{\mathrm{\hat{th}}}(\mathbf{x}) \\
    & \text{subject to}
    & & {g(\mathbf{x})} = 0.8 mm \times V_e \times l_t \le 24 mm^3/s \\
    & 
    & & 215^{\circ}\text{C} \leq T_e \leq 280^{\circ}\text{C} \\
    & 
    & & 25 \text{mm/s} \leq V_e \leq 45 \text{mm/s} \\
    & 
    & & l_t \in \{0.42,0.60,0.70\} \text{mm}\\
    \end{aligned}
\label{eq:bi-obj2}
\end{equation}
Similar to Case 1, the Pareto front obtained for these two objectives is shown in Fig.~\ref{fig:front_2} with selected design points A$_2$, B$_2$, and C$_2$ for experimental validation.

\subsubsection{Case 3}
\begin{figure}[!htb]
\centerline{\includegraphics[width=.85\columnwidth,keepaspectratio]{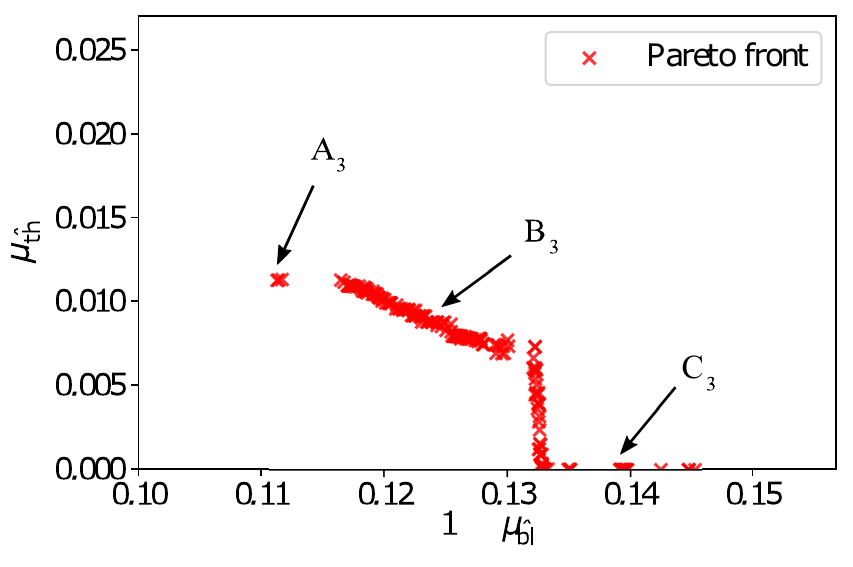}}
	\caption{Pareto front of Case 3: A$_3$ = ($T_e$ ($^{\circ}$C), $V_e$ (mm/s), $l_t$ (mm)) = (217.02, 26.01, 0.42), B$_3$ = (217.05, 27.84, 0.42), C$_3$ = (274.04, 34.29, 0.42)} 
	\label{fig:front_3}
\end{figure}

In this case, one minus the mean of dimensionless bond quality metric ($1-\mu_{\mathrm{\hat{bl}}}$) and the dimensionless mean of part thickness error $\mu_{\mathrm{\hat{th}}}$ are minimized by using both models $\mathrm{BNN}_{\mathrm{bl}}$ and $\mathrm{BNN}_{\mathrm{th}}$. 
\begin{equation}
    \begin{aligned}
    & \underset{\mathbf{x}  \in \mathbb{R}^{n_x}}{\text{minimize}}
    & & w_1\ (1-\mu_{\mathrm{\hat{bl}}}(\mathbf{x})) + w_2\ \mu_{\mathrm{\hat{th}}}(\mathbf{x}) \\
    & \text{subject to}
    & & {g(\mathbf{x})} = 0.8 mm \times V_e \times l_t \le 24 mm^3/s \\
    & 
    & & 215^{\circ}\text{C} \leq T_e \leq 280^{\circ}\text{C} \\
    & 
    & & 25 \text{mm/s} \leq V_e \leq 45 \text{mm/s} \\
    & 
    & & l_t \in \{0.42,0.60,0.70\} \text{mm}\\
    \end{aligned}
\label{eq:bi-obj3}
\end{equation}
The Pareto front of these two objectives is shown in Fig.~\ref{fig:front_3} with selected design points A$_3$, B$_3$, and C$_3$ for experimental validation.

\subsubsection{Case 4}
In this case, one minus the mean of the dimensionless bond quality metric ($\mathrm{Obj}_1 = 1-\mu_{\mathrm{\hat{bl}}}$), the standard deviation of the dimensionless bond quality metric ($\mathrm{Obj}_2=\sigma_{\mathrm{\hat{bl}}}$), and the dimensionless mean and standard deviation of part thickness error ($\mathrm{Obj}_3=\mu_{\mathrm{\hat{th}}}$ \& $\mathrm{Obj}_4=\sigma_{\mathrm{\hat{th}}}$) are minimized by using both models $\mathrm{BNN}_{\mathrm{bl}}$ and $\mathrm{BNN}_{\mathrm{th}}$. In each generation, a new value of the design variables ($T_e$, $V_e$, $l_t$) is generated and passed to the optimizer to evaluate four objectives.
\begin{equation}
    \begin{aligned}
    & \underset{\mathbf{x}  \in \mathbb{R}^{n_x}}{\text{minimize}}
    & & w_1\ (1-\mu_{\mathrm{\hat{bl}}}(\mathbf{x})) + w_2\ \sigma_{\mathrm{\hat{bl}}}(\mathbf{x}) + w_3\ \mu_{\mathrm{\hat{th}}}(\mathbf{x}) + w_4\ \sigma_{\mathrm{\hat{th}}}(\mathbf{x}) \\
    & \text{subject to}
    & & {g(\mathbf{x})} = 0.8 mm \times V_e \times l_t \le 24 mm^3/s \\
    & 
    & & 215^{\circ}\text{C} \leq T_e \leq 280^{\circ}\text{C} \\
    & 
    & & 25 \text{mm/s} \leq V_e \leq 45 \text{mm/s} \\
    & 
    & & l_t \in \{0.42,0.60,0.70\} \text{mm}\\
    \end{aligned}
\label{eq:bi-obj4}
\end{equation}

\begin{figure}[!htb]
\centerline{\includegraphics[width=.85\columnwidth,keepaspectratio]{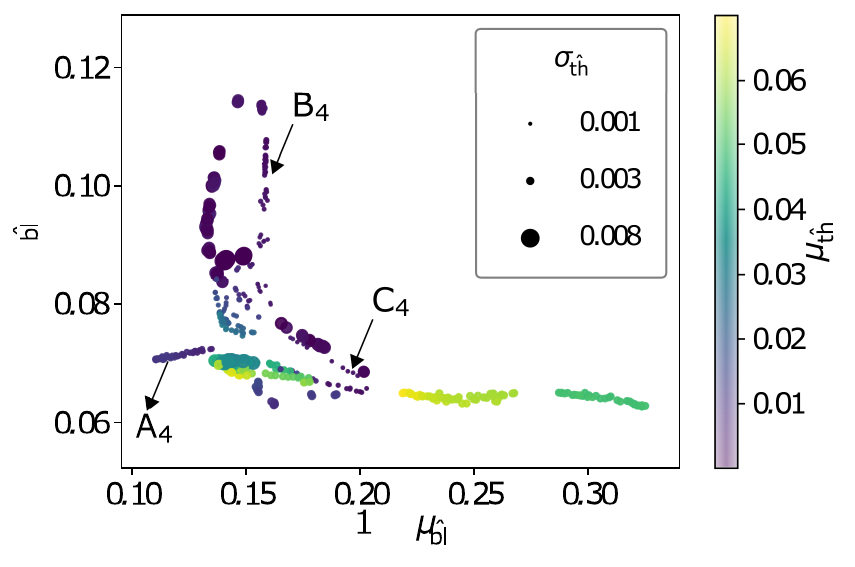}}
	\caption{Pareto designs for Case 4: A$_4$ = ($T_e$ ($^{\circ}$C), $V_e$ (mm/s), $l_t$ (mm)) = (217.08, 26.88, 0.42), B$_4$ = (277.99, 39.41, 0.42), C$_4$ = (225.34, 31.84, 0.42)}
	\label{fig:case4}
\end{figure}

A plot of the Pareto front is demonstrated in Fig.~\ref{fig:case4}, where $\mathrm{Obj}_1$ is displayed on the x-axis, $\mathrm{Obj}_2$ on the y-axis, $\mathrm{Obj}_3$ by the color of the markers, and $\mathrm{Obj}_4$ by the size of the markers. A single design can be selected from the optimal designs shown in Fig.~\ref{fig:case4}. The designer can choose a design based on the relative importance of each objective over the others, since slight improvement in one of the objectives may lead to significant degradation in other objectives. From the optimal solutions, three points A$_4$, B$_4$, and C$_4$ are chosen for experimental validation. 

\begin{figure}[!htb]
\centerline{\includegraphics[width=.85\columnwidth,keepaspectratio]{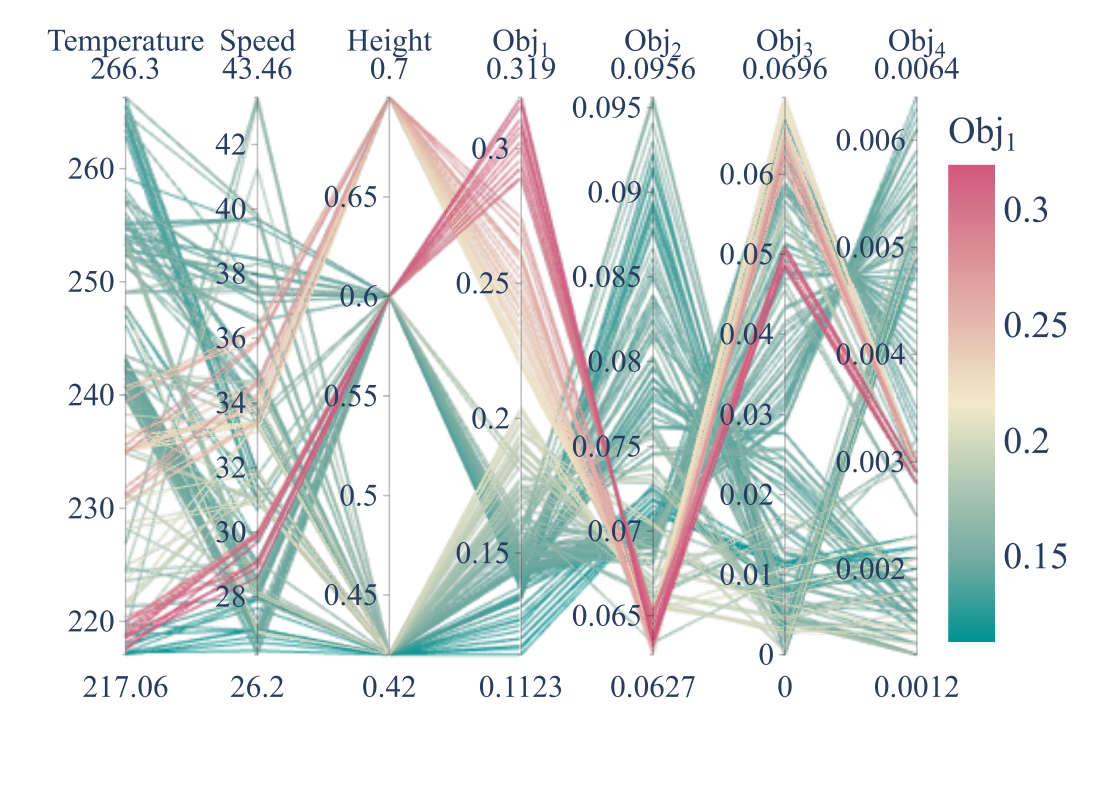}}
	\caption{Parallel coordinate representation of the model dependencies with a color mapping along the first objective ($1-\mu_{\mathrm{\hat{bl}}}$).}
	\label{fig:obj1_parallelplot}
\end{figure}

\begin{figure}[!htb]
\centerline{\includegraphics[width=.85\columnwidth,keepaspectratio]{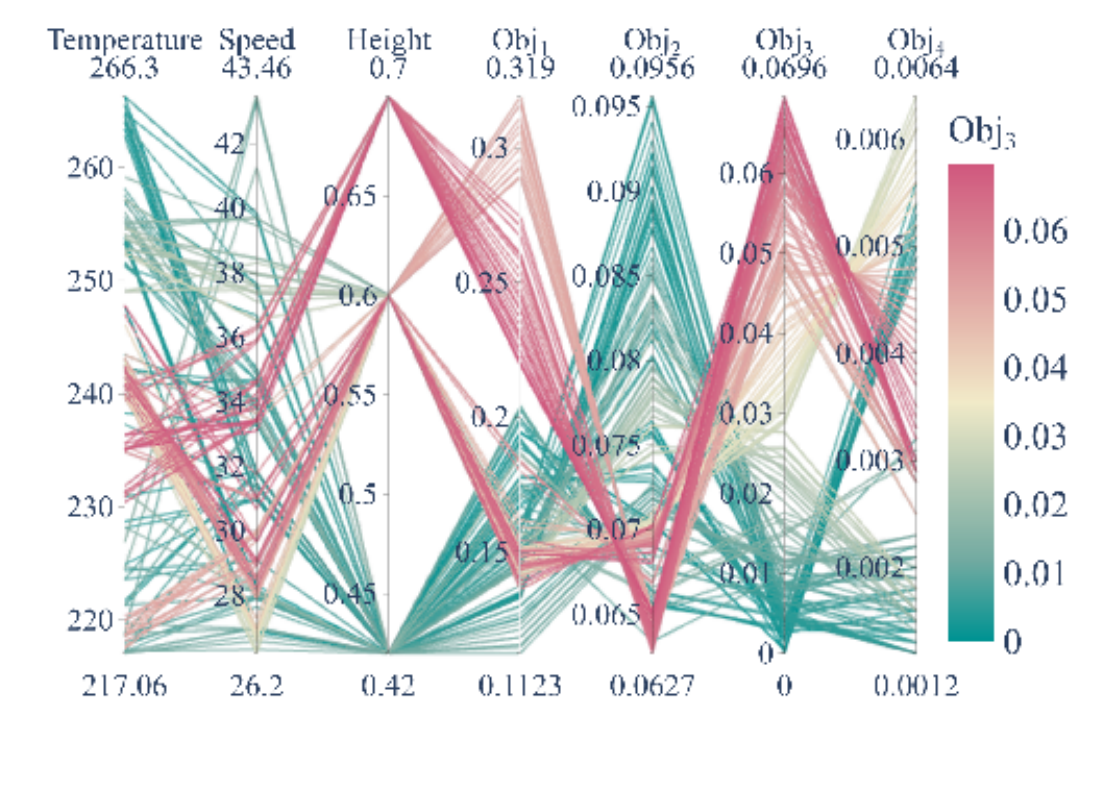}}
	\caption{Parallel coordinate representation of the model dependencies with a color mapping along the third objective ($\mu_{\mathrm{\hat{th}}}$).}
	\label{fig:obj3_parallelplot}
\end{figure}

The optimization results for Case 4 can be illustrated using the parallel coordinate plots shown in Figs.~\ref{fig:obj1_parallelplot} and \ref{fig:obj3_parallelplot}. The resulting parallel coordinates reflect the Pareto dominance relation between different solutions. In Fig.~\ref{fig:obj1_parallelplot}, the color map reflects the values of the first objective ($1-\mu_{\mathrm{\hat{bl}}}$). The red solid line represents the solution that results in the worst mean bond quality within the Pareto front. Note that due to the negative correlation between the first and second objectives (heavily conflicting objectives), worst mean bond quality solutions (i.e., red solid lines) couple with the best (lower) standard deviation bond quality. The mean bond quality starts to improve as the color of the solid line changes from red to green. In the same manner as Fig.~\ref{fig:obj1_parallelplot}, Fig.~\ref{fig:obj3_parallelplot} also illustrates the optimization results with a color map of the third objective. The red solid lines denote the solutions that yield the poorest mean geometric accuracy within the Pareto front approximations. The mean and standard deviation of the geometric accuracy are partially negatively correlated since there are lines that are parallel between the two axes in parallel coordinates. More specifically, the parallel green solid lines at the bottom of the third and fourth objectives that do not intersect with other lines depict the solutions with moderately good mean and standard deviation of geometric accuracy.

\begin{figure}[!htb]
\centerline{\includegraphics[width=.85\columnwidth,keepaspectratio]{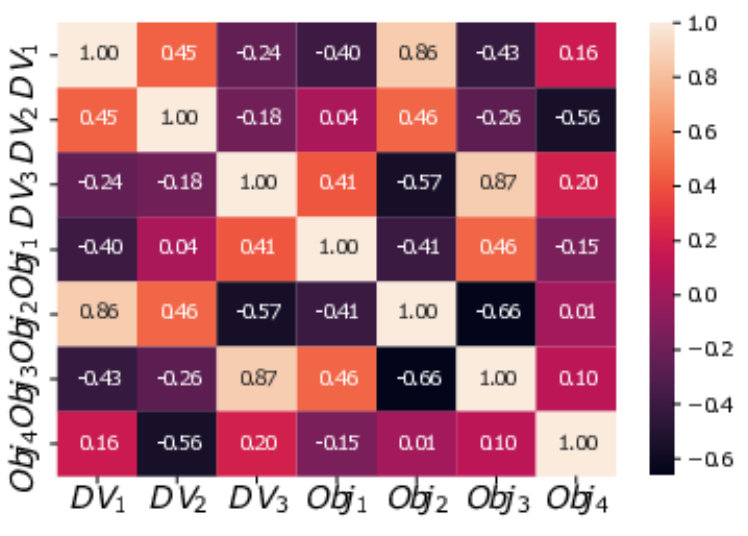}}
	\caption{The correlation coefficient between design variables and objective functions}
	\label{fig:corr}
\end{figure}

The Pearson correlation coefficient between each design variable and/or objective functions is shown in Fig.~\ref{fig:corr}. A correlation coefficient of $\pm 1$ represents a perfect linear relationship. The nozzle temperature (DV$_1$) is negatively correlated with Obj$_1=1-\mu_{\mathrm{\hat{bl}}}$ and Obj$_3=\mu_{\mathrm{\hat{th}}}$, which means that the mean geometric accuracy and the overall mean bond quality of a part tend to improve with increasing nozzle temperature. Whereas, parts with a greater layer height (DV$_3$) appear to have a degraded mean geometrical accuracy and mean bond quality. The nozzle speed (DV$_2$) has a smaller effect on the objective functions. The correlation coefficient between Obj$_1$ and Obj$_3$ is calculated as 0.46 indicating that these two objectives do not have a significant relationship, which is also illustrated in Fig.~\ref{fig:MCS3}. Furthermore, the parallel plots and the correlation matrix demonstrate the difficulty in choosing the optimal process parameters. Therefore, the Pareto front is beneficial in finding a design that offers a good trade-off between the objectives.

The computation time for one optimization is on average ten minutes using a desktop computer (Intel\textsuperscript{\tiny\textregistered} Xeon\textsuperscript{\tiny\textregistered} CPU E5-1660 v4$@$3.20GHz with 32 GB RAM and GPU NVIDIA Quadro K620 with 2 GB).

\subsection{Monte Carlo optimization}

\begin{figure}[!htb]
\centerline{\includegraphics[width=.85\columnwidth,keepaspectratio]{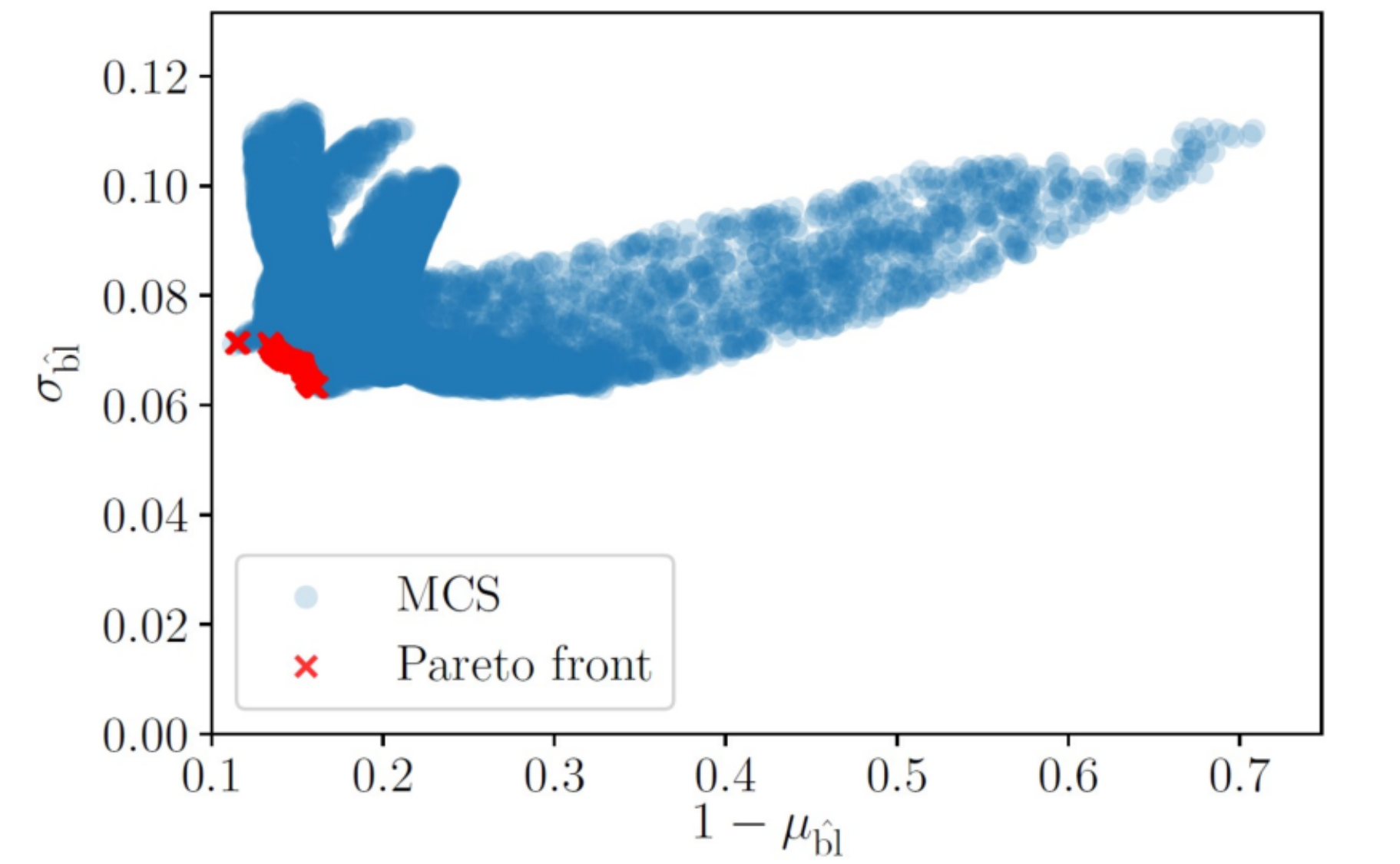}}
	\caption{Pareto front and MCS results of Case 1}
	\label{fig:MCS1}
\end{figure}

\begin{figure}[!htb]
\centerline{\includegraphics[width=.85\columnwidth,keepaspectratio]{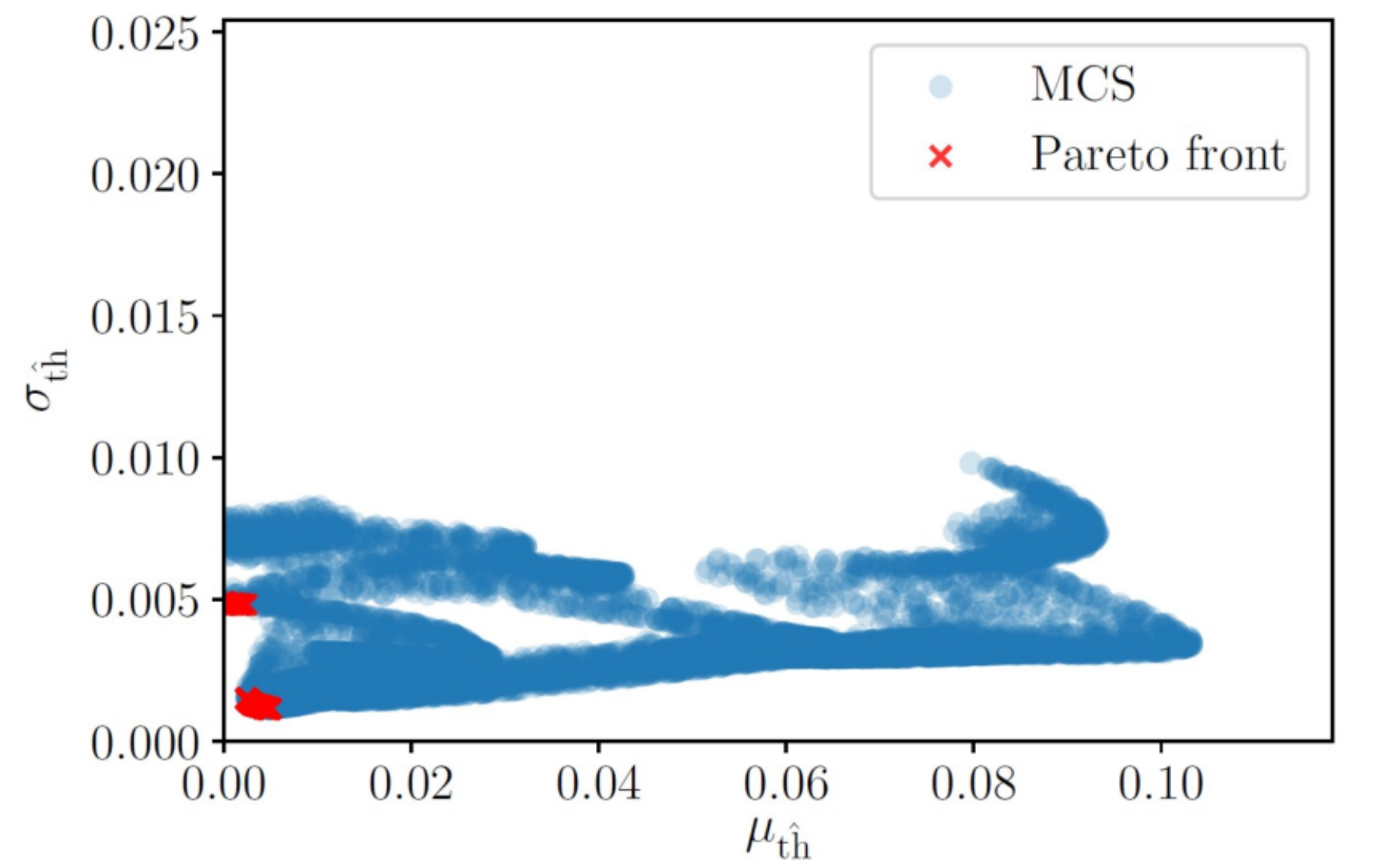}}
	\caption{Pareto front and MCS results of Case 2}
	\label{fig:MCS2}
\end{figure}

\begin{figure}[!htb]
\centerline{\includegraphics[width=.8\columnwidth,keepaspectratio]{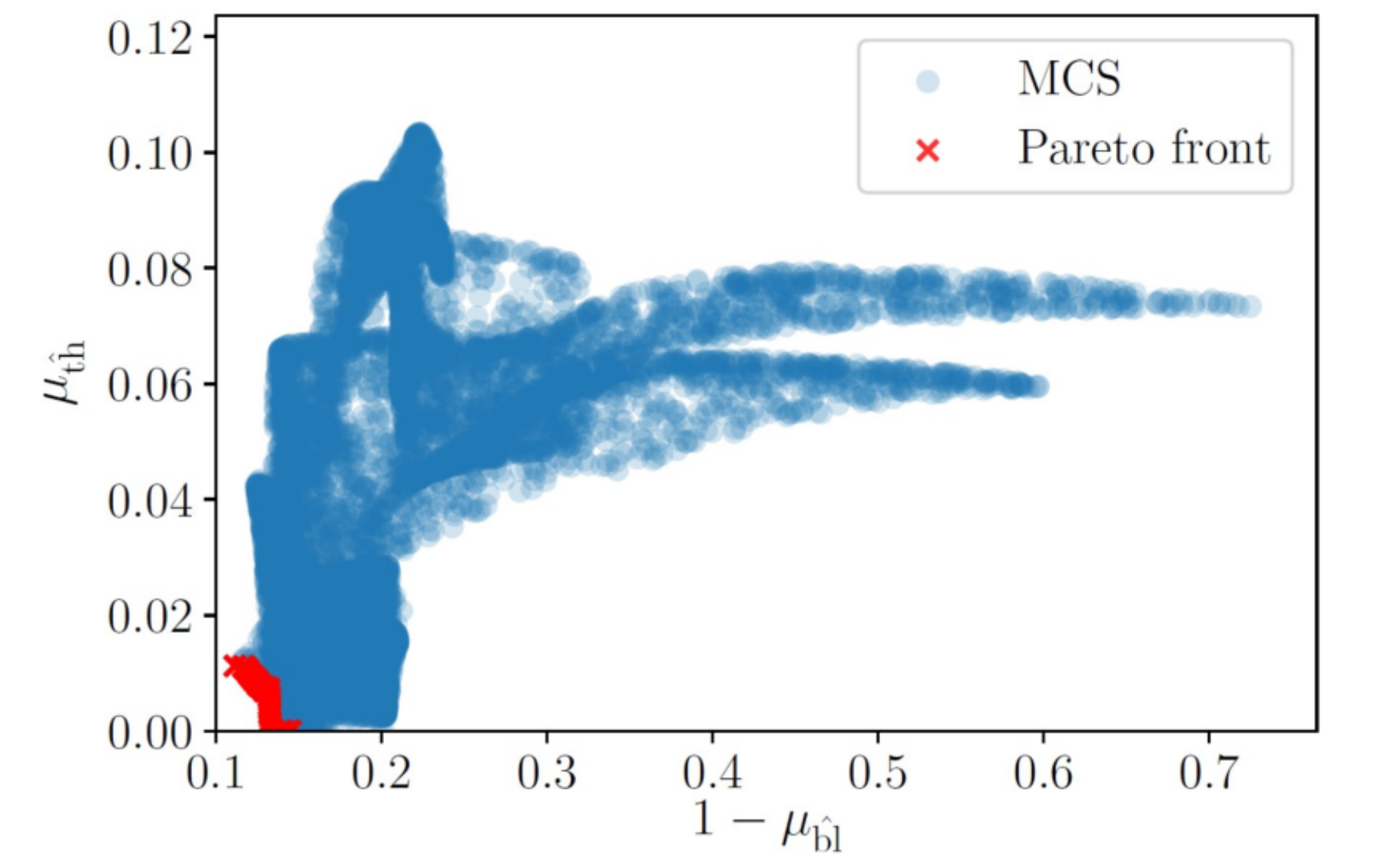}}
	\caption{Pareto front and MCS results of Case 3}
	\label{fig:MCS3}
\end{figure}

The Pareto front and direct correlations between objective functions obtained from the NSGA-II algorithm in Section~\ref{sec:mo_opti} are compared with a Monte Carlo sampling (MCS) approach. A relatively large number of Monte Carlo samples (10,000 samples) of design variables from the design space are generated and the objective functions for different cases are predicted using the trained BNN models. Due to computational complexity and difficulty in visualizing the four-dimensional results of Case 4, MCS approach is only employed for bi-objective cases i.e., Cases 1, 2 and 3 (Figs.~\ref{fig:MCS1},~\ref{fig:MCS2},~\ref{fig:MCS3}).

The relationship between the two objectives in Case 2 and 3 as shown in Figs.~\ref{fig:MCS2}~and \ref{fig:MCS3} is highly non-linear. The regions with sharp edges generally represent constraint boundaries. The relationship between two physical quantities, i.e., bond quality and geometry accuracy of a part, shown in Fig.~\ref{fig:MCS3} demonstrates the importance of choosing the optimal process parameters. For some combinations of design variables, both physical quantities may have significantly degraded values, which can be avoided by using the proposed methodology. The Pareto front results shown in Figs.~\ref{fig:front_1},~\ref{fig:front_2} and \ref{fig:front_3} are superimposed on the MCS results and labeled as red crosses in Figs.~\ref{fig:MCS1},~\ref{fig:MCS2}, and~\ref{fig:MCS3}. The superimposed Pareto fronts show that the optimization results are in agreement with the MCS results.

\subsection{Experimental validation}
Three points are chosen for validation from the Pareto fronts for each of the four multi-objective optimization cases considered above. The predicted values of objectives for each point are validated using classical hypothesis testing. The sample size of experiments for each point is 5. The sample size of model predictions for each point is 100, which is produced using MC dropout. For validation of the prediction models, the $t$-test, which is based on Student's $t$-distribution, is used to test the hypothesis that the mean of the observations is equal to the mean of the model prediction. Since the calculated $p$-values for all selected design points (A$_1$, B$_1$, C$_1$, A$_2$, B$_2$, C$_2$, A$_3$, B$_3$, C$_3$, A$_4$, B$_4$, C$_4$) are above the significance level $\alpha=0.05$, the predictions at the optimum solutions agree with the experimental observations. This is the validation of the BNN prediction models, which are trained with the previous experimental data. Note that the training data and validation data are separate. 

\begin{figure}[!htb]
\centerline{\includegraphics[width=.85\columnwidth,keepaspectratio]{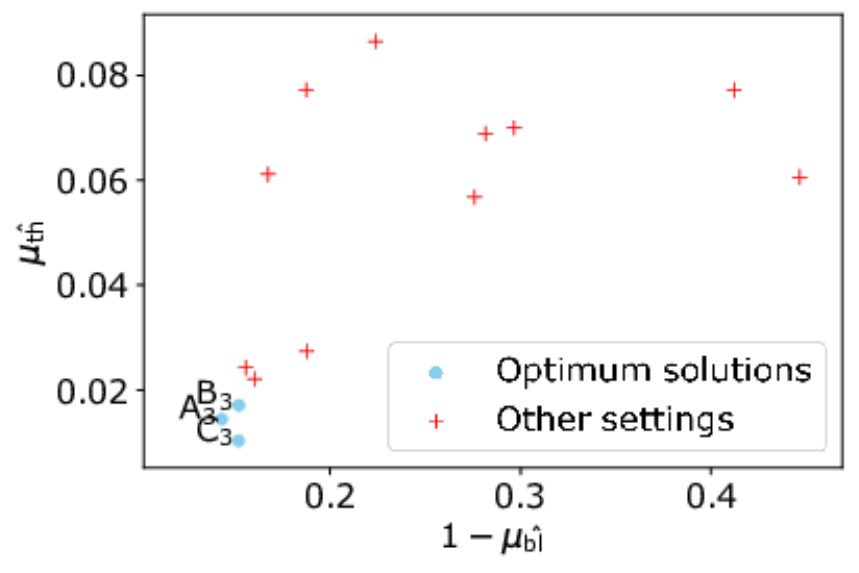}}
	\caption{Experimental validation of optimal process design (Case 3)}
	\label{fig:OptSol}
\end{figure}

Next we also validate the optimization result, by demonstrating that the optimum solution gives better part quality than randomly selected process parameter settings. Parts are printed and measured at eleven other process parameter settings selected at random and compared to the part printed with the optimal process parameters of Case 3. The dimensionless mean thickness error ($\mu_{\mathrm{\hat{th}}}$) and the
dimensionless bond quality metric ($1-\mu_{\mathrm{\hat{bl}}}$) obtained using the randomly selected process parameters are found to be larger than that of the optimal solution (Fig.~\ref{fig:OptSol}), thus clearly demonstrating the effectiveness of the proposed methodology.

\subsection{Summary of results}

We find that, using the proposed approach, we were able to manufacture FFF parts with better part quality using the optimization results from the Pareto front obtained through the probabilistic data-driven methodology for process parameter optimization. In this study, we obtain probabilistic estimates of both quantities (bond quality and geometrical accuracy) and include the probabilistic estimates within the process design for optimization under uncertainty. The input variability in the experimental data and the model uncertainty are captured through the probabilistic deep learning approach, BNN. 

The probabilistic estimates facilitate different options of optimization under uncertainty: optimize only the mean value of an individual quality metric, optimize both the mean and variance of an individual quality metric, optimize the mean values of multiple quality metrics simultaneously, and optimize both the mean values and variances of multiple quality metrics simultaneously. Except the first option, all other options require multi-objective formulations. It is observed that the mean values of both quality metrics (bond length and part thickness accuracy) are optimized with the smallest layer thickness (i.e., 0.42 mm). This is generally expected in AM processes. However, it is not that straightforward to see a similar trend for other parameters. In general, the optimal solutions for the mean of both quality metrics are achieved with larger nozzle temperature and smaller nozzle speed values. However, there are some cases where some other combinations of nozzle temperature and speed (e.g., small temperature and speed) also result in optimal solutions for these two quantities. Thus, the Pareto front is helpful  in finding a design that offers a good trade-off between the objectives.

The prediction models are verified using the cross-validation approach as explained in Section~\ref{sec:crossval}, and validated against experimental data. The optimization results are validated by printing and measuring the parts corresponding to the points A, B, and C of the Pareto fronts. The results show that the optimum solution using the proposed process optimization framework gives better part quality than randomly selected process parameter settings.

\section{Conclusion}\label{sec:conclusion}
In this paper, a data-driven machine learning model, is proposed for multi-objective optimization under uncertainty of process parameters in AM. The proposed methodology is demonstrated for AM by optimizing the FFF process parameters considering the objectives of minimizing the error in the part geometry and maximizing the bond quality between the filaments. Several cases with multiple objectives are considered: minimizing the mean and standard deviation of the bond quality metric, minimizing the mean and standard deviation of part thickness error, minimizing the mean part thickness error and mean bond quality metric, and minimizing the means and standard deviations of both part thickness error and bond quality metric. The BNN models were constructed using actual manufacturing data, and are able to quantify the uncertainty in the prediction. Both the BNN models and the optimization solutions are validated with actual manufactured parts, at the optimal and various other settings. The parts manufactured with optimal process parameter settings are observed to result in better part quality than parts manufactured with other non-optimal settings. 

In future work, the proposed methodology can be extended to process control in AM. Using the real-time data collected during the manufacturing process, the approach can be extended to make layer-wise control decisions under uncertainty about the AM process such as changing the process parameters for subsequent layers to improve the part by monitoring the quality metrics during manufacturing. Future work can also incorporate the physics knowledge to fill the physics knowledge gap in the machine learning model, and to improve the prediction accuracy by correcting for the approximations in the physics-based models.

\section*{Acknowledgment}
This study was supported by funding from the National Institute of Standards and Technology under the Smart Manufacturing Data Analytics Project (Cooperative Agreement No. 70NANB14H036). The support is gratefully acknowledged. The authors also thank Joseph D. Olson, research assistant at Vanderbilt University, for conducting some of the experiments. \\
\\
\textbf{Disclaimer}\\
Certain commercial systems are identified in this paper. Such identification does not imply recommendation or endorsement by NIST; nor does it imply that the products identified are necessarily the best available for the purpose. Further, any opinions, findings, conclusions, or recommendations expressed in this material are those of the authors and do not necessarily reflect the views of NIST or any other supporting U.S. government or corporate organizations.

\bibliographystyle{unsrt}  
\bibliography{references}

@article{wang2019data,
  title={A Data-Driven Approach for Process Optimization of Metallic Additive Manufacturing Under Uncertainty},
  author={Wang, Zhuo and Liu, Pengwei and Xiao, Yaohong and Cui, Xiangyang and Hu, Zhen and Chen, Lei},
  journal={Journal of Manufacturing Science and Engineering},
  volume={141},
  number={8},
  year={2019},
  doi={10.1115/1.4043798},
  publisher={American Society of Mechanical Engineers Digital Collection}
}

@article{nath2020optimization,
title = "Optimization of fused filament fabrication process parameters under uncertainty to maximize part geometry accuracy",
journal = "Additive Manufacturing",
volume = "35",
pages = "101331",
year = "2020",
issn = "2214-8604",
doi = "10.1016/j.addma.2020.101331",
author = "Paromita Nath and Joseph D. Olson and Sankaran Mahadevan and Yung-Tsun Tina Lee"
}

@article{mahmoudi2018multivariate,
  title={Multivariate calibration and experimental validation of a 3{D} finite element thermal model for laser powder bed fusion metal additive manufacturing},
  author={Mahmoudi, Mohamad and Tapia, Gustavo and Karayagiz, Kubra and Franco, Brian and Ma, Ji and Arroyave, Raymundo and Karaman, Ibrahim and Elwany, Alaa},
  journal={Integrating Materials and Manufacturing Innovation},
  volume={7},
  number={3},
  pages={116--135},
  year={2018},
  doi={10.1007/s40192-018-0113-z},
  publisher={Springer}
}

@misc{international2015additive,
  title={{ISO/ASTM} 52900:2015 {A}dditive manufacturing - General principles - terminology},
  author={{ASTM Standard}},
  year={2015},
  publisher={Geneva, Switzerland}
}

@article{megahed2016metal,
  title={Metal additive-manufacturing process and residual stress modeling},
  author={Megahed, Mustafa and Mindt, Hans-Wilfried and N’Dri, Narcisse and Duan, Hongzhi and Desmaison, Olivier},
  journal={Integrating Materials and Manufacturing Innovation},
  volume={5},
  number={1},
  pages={61--93},
  year={2016},
  doi={10.1186/s40192-016-0047-2},
  publisher={Springer}
}

@article{baumann2018trends,
  title={Trends of machine learning in additive manufacturing},
  author={Baumann, Felix W and Sekulla, Andr{\'e} and Hassler, Michael and Himpel, Benjamin and Pfeil, Markus},
  journal={International Journal of Rapid Manufacturing},
  volume={7},
  number={4},
  pages={310--336},
  year={2018},
  doi={10.1504/IJRAPIDM.2018.095788},
  publisher={Inderscience Publishers (IEL)}
}

@article{debroy2017building,
  title={Building digital twins of 3D printing machines},
  author={Debroy, Tarasankar and Zhang, Wei and Turner, J and Babu, Sudarsanam Suresh},
  journal={Scripta Materialia},
  volume={135},
  pages={119--124},
  year={2017},
  doi={10.1016/j.scriptamat.2016.12.005},
  publisher={Elsevier}
}

@article{nath2019uncertainty,
  title={Uncertainty quantification of grain morphology in laser direct metal deposition},
  author={Nath, Paromita and Hu, Zhen and Mahadevan, Sankaran},
  journal={Modelling and Simulation in Materials Science and Engineering},
  volume={27},
  number={4},
  pages={044003},
  year={2019},
  doi={10.1088/1361-651X/ab1676},
  publisher={IOP Publishing}
}

@article{gaikwad2019situ,
  title={In Situ Monitoring of Thin-Wall Build Quality in Laser Powder Bed Fusion Using Deep Learning},
  author={Gaikwad, Aniruddha and Imani, Farhad and Yang, Hui and Reutzel, Edward and Rao, Prahalada},
  year={2019},
  volume = "3",
  doi={10.1520/SSMS20190027},
  publisher={ASTM International}
}

@article{ding2016bead,
  title={Bead modelling and implementation of adaptive MAT path in wire and arc additive manufacturing},
  author={Ding, Donghong and Pan, Zengxi and Cuiuri, Dominic and Li, Huijun and van Duin, Stephen and Larkin, Nathan},
  journal={Robotics and Computer-Integrated Manufacturing},
  volume={39},
  pages={32--42},
  year={2016},
  doi={10.1016/j.rcim.2015.12.004},
  publisher={Elsevier}
}

@article{mehrpouya2019prediction,
  title={A prediction model for finding the optimal laser parameters in additive manufacturing of NiTi shape memory alloy},
  author={Mehrpouya, Mehrshad and Gisario, Annamaria and Rahimzadeh, Atabak and Nematollahi, Mohammadreza and Baghbaderani, Keyvan Safaei and Elahinia, Mohammad},
  journal={The International Journal of Advanced Manufacturing Technology},
  volume={105},
  number={11},
  pages={4691--4699},
  year={2019},
  doi={10.1007/s00170-019-04596-z},
  publisher={Springer}
}

@article{ye2018situ,
  title={In situ monitoring of selective laser melting using plume and spatter signatures by deep belief networks},
  author={Ye, Dongsen and Fuh, Jerry Ying Hsi and Zhang, Yingjie and Hong, Geok Soon and Zhu, Kunpeng},
  journal={ISA transactions},
  volume={81},
  pages={96--104},
  year={2018},
  doi={10.1016/j.isatra.2018.07.021},
  publisher={Elsevier}
}

@article{kwon2020convolutional,
  title={A convolutional neural network for prediction of laser power using melt-pool images in laser powder bed fusion},
  author={Kwon, Ohyung and Kim, Hyung Giun and Kim, Wonrae and Kim, Gun-Hee and Kim, Kangil},
  journal={IEEE Access},
  year={2020},
  doi={10.1109/ACCESS.2020.2970026},
  publisher={IEEE}
}

@article{khanzadeh2018porosity,
  title={Porosity prediction: Supervised-learning of thermal history for direct laser deposition},
  author={Khanzadeh, Mojtaba and Chowdhury, Sudipta and Marufuzzaman, Mohammad and Tschopp, Mark A and Bian, Linkan},
  journal={Journal of manufacturing systems},
  volume={47},
  pages={69--82},
  year={2018},
  doi={10.1016/j.jmsy.2018.04.001},
  publisher={Elsevier}
}

@article{schneider2012nih,
  title={NIH Image to ImageJ: 25 years of image analysis},
  author={Schneider, Caroline A and Rasband, Wayne S and Eliceiri, Kevin W},
  journal={Nature methods},
  volume={9},
  number={7},
  pages={671},
  year={2012},
  publisher={Nature Publishing Group}
}

@article{deb2002fast,
  title={A fast and elitist multiobjective genetic algorithm: NSGA-II},
  author={Deb, Kalyanmoy and Pratap, Amrit and Agarwal, Sameer and Meyarivan, TAMT},
  journal={IEEE transactions on evolutionary computation},
  volume={6},
  number={2},
  pages={182--197},
  year={2002},
  doi={10.1109/4235.996017},
  publisher={IEEE}
}

@InProceedings{gal2016dropout,
  title = 	 {Dropout as a Bayesian Approximation: Representing Model Uncertainty in Deep Learning},
  author = 	 {Yarin Gal and Zoubin Ghahramani},
  booktitle = 	 {Proceedings of The 33rd International Conference on Machine Learning},
  pages = 	 {1050--1059},
  year = 	 {2016},
  volume = 	 {48},
  series = 	 {Proceedings of Machine Learning Research},
  address = 	 {New York, New York, USA},
  month = 	 {20--22 Jun},
  publisher =    {PMLR},
  url = 	 {http://proceedings.mlr.press/v48/gal16.html}
}

@inproceedings{kendall2017uncertainties,
 author = {Kendall, Alex and Gal, Yarin},
 booktitle = {Proceedings of the 31st International Conference on Neural Information Processing Systems},
 publisher = {Curran Associates, Inc.},
 title = {What Uncertainties Do We Need in Bayesian Deep Learning for Computer Vision?},
 volume = {30},
 isbn = {9781510860964},
 pages = {5580–5590},
 series = {NIPS'17},
 year = {2017}
}

@article{du2004sequential,
  title={Sequential optimization and reliability assessment method for efficient probabilistic design},
  author={Du, Xiaoping and Chen, Wei},
  journal={Journal of mechanical design},
  volume={126},
  number={2},
  pages={225--233},
  year={2004},
  doi={10.1115/1.1649968},
  publisher={American Society of Mechanical Engineers Digital Collection}
}

@article{zaman2011robustness,
  title={Robustness-based design optimization under data uncertainty},
  author={Zaman, Kais and McDonald, Mark and Mahadevan, Sankaran and Green, Lawrence},
  journal={Structural and Multidisciplinary Optimization},
  volume={44},
  number={2},
  pages={183--197},
  year={2011},
  doi={10.1007/s00158-011-0622-2},
  publisher={Springer}
}

@InProceedings{blundell2015weight,
  title = 	 {Weight Uncertainty in Neural Network},
  author = 	 {Charles Blundell and Julien Cornebise and Koray Kavukcuoglu and Daan Wierstra},
  booktitle = {Proceedings of the 32nd International Conference on Machine Learning},
  pages = 	 {1613--1622},
  year = 	 {2015},
  volume = 	 {37},
  publisher =    {PMLR},
  url = 	 {http://proceedings.mlr.press/v37/blundell15.html}
}

@inproceedings{denker1991transforming,
  title={Transforming neural-net output levels to probability distributions},
  author={Denker, John S and LeCun, Yann},
  booktitle={Advances in neural information processing systems},
  pages={853--859},
  year={1991}
}

@article{mackay1992practical,
  title={A Practical Bayesian Framework for Backpropagation Networks},
  author={MacKay, David JC},
  journal={Neural Computation},
  volume={4},
  number={3},
  pages={448--472},
  year={1992},
  publisher={MIT Press}
}

@article{neal2012bayesian,
author="Neal, R. M.",
title="Bayesian Learning for Neural Networks",
journal="Lecture Notes in Statistics",
ISSN="",
publisher="Springer",
year="1996",
month="",
volume="",
number="",
pages="",
URL="https://ci.nii.ac.jp/naid/10017595601/en/",
DOI="",
}

@article{knowles2006tutorial,
  title={A tutorial on the performance assessment of stochastic multiobjective optimizers},
  author={Knowles, Joshua D and Thiele, Lothar and Zitzler, Eckart},
  journal={TIK-Report},
  volume={214},
  year={2006},
  doi={10.3929/ethz-b-000023822},
  publisher={ETH Zurich}
}

@inproceedings{zitzler2007hypervolume,
  title={The hypervolume indicator revisited: On the design of Pareto-compliant indicators via weighted integration},
  author={Zitzler, Eckart and Brockhoff, Dimo and Thiele, Lothar},
  booktitle={International Conference on Evolutionary Multi-Criterion Optimization},
  pages={862--876},
  year={2007},
  doi={10.1007/978-3-540-70928-2_64},
  organization={Springer}
}

@article{zitzler2000comparison,
  title={Comparison of multiobjective evolutionary algorithms: Empirical results},
  author={Zitzler, Eckart and Deb, Kalyanmoy and Thiele, Lothar},
  journal={Evolutionary computation},
  volume={8},
  number={2},
  pages={173--195},
  year={2000},
  doi={https://doi.org/10.1162/106365600568202},
  publisher={MIT Press}
}

@article{zitzler2003performance,
  title={Performance assessment of multiobjective optimizers: An analysis and review},
  author={Zitzler, Eckart and Thiele, Lothar and Laumanns, Marco and Fonseca, Carlos M and Da Fonseca, Viviane Grunert},
  journal={IEEE Transactions on evolutionary computation},
  volume={7},
  number={2},
  pages={117--132},
  year={2003},
  publisher={IEEE}
}

@article{berkcan2020Optimization,
    title = "Process Optimization under Uncertainty for Improving the Bond Quality of Polymer Filaments in Fused Filament Fabrication",
    journal = "Journal of Manufacturing Science and Engineering",
    pages = "1-46",
    year = "2020",
    volume = "143",
    issue = "2",
    month = "08",
    issn = "1087-1357",
    doi = "10.1115/1.4048073",
    author = "Kapusuzoglu, Berkcan and Sato, Matthew and Mahadevan, Sankaran and Witherell, Paul"
}

@article{berkcan2020PIML,
  title={Physics-Informed and Hybrid Machine Learning in Additive Manufacturing: Application to Fused Filament Fabrication},
  author={Kapusuzoglu, Berkcan and Mahadevan, Sankaran},
  journal={JOM},
  pages={4695--4705},
  volume = {72},
  year={2020},
  doi={10.1007/s11837-020-04438-4},
  publisher={Springer}
}

@MISC{ultimakerS5,
    title = {Ultimaker {S}5 {S}pecifications},
    year = {accessed: 02/01/2020},
    howpublished={\url{https://ultimaker.com/3d-printers/ultimaker-s5}}
}

@MISC{LKsensor,
    title = {Keyence {LK-H}057 {S}pecifications},
    year = {accessed: 07/29/2019},
    howpublished={\url{https://www.keyence.com/products/measure/laser-1d/lk-g5000/models/lk-h057/index.jsp}}
}

@article{ling2013quantitative,
  title={Quantitative model validation techniques: New insights},
  author={Ling, You and Mahadevan, Sankaran},
  journal={Reliability Engineering \& System Safety},
  volume={111},
  pages={217--231},
  year={2013},
  publisher={Elsevier}
}

@article{gilks2005m,
  title={Markov Chain Monte Carlo},
  author={Gilks, Walter R},
  journal={Encyclopedia of Biostatistics},
  volume={4},
  year={2005},
  doi={10.1002/0470011815.b2a14021},
  publisher={Wiley Online Library}
}

@article{arulampalam2002tutorial,
  title={A tutorial on particle filters for online nonlinear/non-Gaussian Bayesian tracking},
  author={Arulampalam, M Sanjeev and Maskell, Simon and Gordon, Neil and Clapp, Tim},
  journal={IEEE Transactions on signal processing},
  volume={50},
  number={2},
  doi={10.1109/78.978374},
  pages={174--188},
  year={2002},
  publisher={Ieee}
}

@article{beaumont2002approximate,
  title={Approximate Bayesian computation in population genetics},
  author={Beaumont, Mark A and Zhang, Wenyang and Balding, David J},
  journal={Genetics},
  volume={162},
  number={4},
  pages={2025--2035},
  year={2002},
  doi={10.1371/journal.pcbi.1002803},
  publisher={Genetics Soc America}
}

@article{kapusuzoglu2021information,
  title={Information fusion and machine learning for sensitivity analysis using physics knowledge and experimental data},
  author={Kapusuzoglu, Berkcan and Mahadevan, Sankaran},
  journal={Reliability Engineering \& System Safety},
  volume={214},
  pages={107712},
  year={2021},
  doi={10.1016/j.ress.2021.107712},
  publisher={Elsevier}
}

@article{levine2020outcomes,
  title={Outcomes and conclusions from the 2018 AM-Bench measurements, challenge problems, modeling submissions, and conference},
  author={Levine, Lyle and Lane, Brandon and Heigel, Jarred and Migler, Kalman and Stoudt, Mark and Phan, Thien and Ricker, Richard and Strantza, Maria and Hill, Michael and Zhang, Fan and others},
  journal={Integrating Materials and Manufacturing Innovation},
  volume={9},
  number={1},
  pages={1--15},
  year={2020},
  doi={10.1007/s40192-019-00164-1},
  publisher={Springer}
}

@article{lane2020measurements,
  title={Measurements of melt pool geometry and cooling rates of individual laser traces on IN625 bare plates},
  author={Lane, Brandon and Heigel, Jarred and Ricker, Richard and Zhirnov, Ivan and Khromschenko, Vladimir and Weaver, Jordan and Phan, Thien and Stoudt, Mark and Mekhontsev, Sergey and Levine, Lyle},
  journal={Integrating Materials and Manufacturing Innovation},
  pages={1--15},
  year={2020},
  doi={10.1007/s40192-020-00169-1},
  publisher={Springer}
}

@article{cole2020amb2018,
  title={AMB2018-03: Benchmark Physical Property Measurements for Material Extrusion Additive Manufacturing of Polycarbonate},
  author={Cole, Daniel P and Gardea, Frank and Henry, Todd C and Seppala, Jonathan E and Garboczi, Edward J and Migler, Kalman D and Shumeyko, Christopher M and Westrich, Jeffrey R and Orski, Sara V and Gair, Jeffrey L},
  journal={Integrating Materials and Manufacturing Innovation},
  volume={9},
  number={4},
  pages={358--375},
  year={2020},
  doi={10.1007/s40192-020-00188-y},
  publisher={Springer}
}

\end{document}